\documentclass[useAMS,usenatbib,usegraphicx]{mn2e}

\usepackage{color, aas_macros}
\def\eref#1{equation (\ref{eq:#1})}
\def\eeref#1{(\ref{eq:#1})}
\def\fref#1{Fig. \ref{fig:#1}}
\def\sref#1{section \ref{sec:#1}}

\def\change#1{{#1}}
\def\cchange#1{{#1}}

\title[Detectability of gravitational waves from magnetic mountains]{Improved estimate of the detectability of gravitational radiation
  from a magnetically confined mountain on an accreting neutron star}

\author[M. Vigelius et al.]{M.~Vigelius $^1$\thanks{E-mail: mvigeliu@physics.unimelb.edu.au} and A.~Melatos $^1$ \\
 $^1$ School of Physics, University of Melbourne, Parkville, VIC
 3010, Australia}

\begin{document}

\date{Submitted to MNRAS}

\maketitle

\begin{abstract}
  We give an improved estimate of the detectability of gravitational
  waves from magnetically confined mountains on accreting neutron
  stars. The improved estimate includes the following effects for the
  first time: three-dimensional hydromagnetic (``fast'') relaxation,
  three-dimensional resistive (``slow'') relaxation,   realistic
  accreted masses  $M_a 
  \la 2 \times 10^{-3} M_\odot$,   (where the mountain is grown ab
  initio by injection), and verification of the curvature rescaling
  transformation employed in previous work. Typically, a mountain 
  does not relax appreciably over the lifetime
  of a low-mass X-ray binary. The ellipticity reaches
  $\epsilon \approx 2 \times 10^{-5}$ for $M_a=2\times 10^{-3}
  M_\odot$. The gravitational wave spectrum for triaxial equilibria
  contains an additional line, which, although weak,
  provides valuable information about the mountain shape. We evaluate
  the detectability of magnetic mountains with Initial and Advanced
  LIGO. For a standard, coherent matched filter search, we find a
  signal-to-noise ratio of $d = 28 (M_a/10^{-4} M_\odot) 
  (1+5.5 M_a/10^{-4} M_\odot)^{-1} (D/10\,\mathrm{kpc})^{-1} (T_0/14\,
  \mathrm{d})^{1/2}$ for Initial LIGO, where $D$ is the distance and
  $T_0$ is the observation time. From the nondetection of gravitational
  waves from low-mass X-ray
  binaries to date, and the wave strain limits implied by the spin
  frequency distribution of these objects (due to  gravitational wave
  braking), we conclude that there are other, as yet unmodelled,
  physical effects that further reduce the quadrupole moment of a
  magnetic mountain, most notably sinking into the crust.
\end{abstract}

\bibliographystyle{mn2e}


\section{Introduction}
Accreting neutron stars in low-mass X-ray binaries (LMXBs) are promising
sources of continuous gravitational waves (GWs). The signal from these
emitters can be coherently integrated, so that the signal-to-noise
ratio increases with the square root of the observation time \citep{JKSI}.
Recent directed searches for GWs from the nearby X-ray source Sco
X$-$1 by the Laser Interferometer Gravitational Wave Observatory (LIGO)
set an upper bound on the gravitational wave strain of $h_0 \la
10^{-22}$ \citep{LIGO06}.

LMXBs emit continuous gravitational waves via a variety of physical
mechanisms \citep{Owen06, LIGO06}: nonaxisymmetric
elastic deformations of the neutron star crust, generated by
temperature gradients \citep{Bildsten98, Ushomirsky00, Haskell07} or internal toroidal
magnetic fields \citep{Cutler02}; r-modes, generated by the Chandrasekhar-Friedmann-Schutz
instability \citep{Owen98, Andersson99, Ster03, Nayyar06}; free
precession, excited by internal or accretion torques \citep{Jones02,
  vandenbroeck05, Payne06a, chung08}; and magnetically confined mountains
\citep{Payne04, Melatos05, Payne06a, Vigelius08a, Vigelius08b}.

In the latter mechanism, accreting plasma accumulates at
the magnetic poles and spreads equatorwards. The
frozen-in magnetic field is carried along with the spreading plasma
and is therefore compressed, to the point where magnetic tension
counterbalances the latitudinal pressure gradient. This equilibrium
configuration is termed a magnetic mountain \citep{Payne04}. During
the process, the magnetic dipole moment of the star decreases with
accreted mass, consistent with observational data \citep{Zhang98, Melatos01,
  Payne04, Zhang06}. The distorted magnetic field can also act as a
thermal barrier between the hemispheres, affecting the physics of type
I X-ray bursts \citep{Payne06b}.

In this paper, we draw together the latest analytic and numerical
modelling of magnetic mountains in LMXBs \citep{Vigelius08a,
  Vigelius08b} to compute rigorously the signal-to-noise ratio of
these sources for a coherent search with LIGO. \change{To this end, we
make extensive use of previously published results. Hydromagnetic
equilibria of magnetically confined mountains were computed
analytically and numerically by \citet{Payne04}. While these
configurations are stable to axisymmetric perturbations
\citep{Payne07}, the (nonaxisymmetric) undulating submode of
the three-dimensional Parker instability induces a reconfiguration of
the magnetic field \citep{Vigelius08a}. However,
the line-tying boundary condition at the
stellar surface prevents the mountain from being disrupted and the
saturation state of the instability still exhibits a substantial
quadrupole moment with a high degree of axisymmetry. \citet{Vigelius08b} extend the
analysis to include resistive effects and demonstrate that
the mountain decays on the diffusion time scale. No
evidence for resistive instabilities that grow on a short time scale
is found.}

\change{In their stability analysis, \citet{Payne06a} and
  \citet{Vigelius08a, Vigelius08b} numerically computed an
  equilibrium configuration with a particular value for the accreted
  mass, $M_\mathrm{a}$, subsequently
  loaded this equilibrium into a magnetohydrodynamic (MHD) solver and
  evolved it. In this article, we solve the initial-value MHD problem
  for the first time by injecting plasma into an initially dipolar
  field. This method allows us to independently validate and extend
  the scalings given by \citet{Payne04} and \citet{Melatos05}, which
  were calculated analytically in the small-$M_\mathrm{a}$
  approximation. Taking into account resistive effects and
  three-dimensional reconfiguration of the mountain, we present easily
  applicable formulas to compute the mass ellipticity for a given
  $M_\mathrm{a}$ and give improved estimates on the strength of the
  gravitational wave emission. In particular, we investigate how the
  (small) degree of nonaxisymmetry changes the gravitational wave
  spectrum and how gravitational-wave spectrometry can be used to
  obtain valuable information about the underlying field
  configuration. This analysis ties in with \citet{Payne06a} who
  consider axisymmetric mountains.}

The paper is organised
as follows. We describe quantitatively the physics of the mass
quadrupole moment of a magnetic mountain in section
\ref{sec:mountains}, including three-dimensional force balance, slow
(resistive) and fast (hydromagnetic) relaxation, realistic accreted
masses, and the influence of stellar curvature. Taking into account
these effects, we give a recipe to compute the quadrupole moment as a
function of accreted $M_a$  in section \ref{sec:gravwaves} and hence estimate
the strength and detectability of the GW signal. The frequency
spectrum of the signal is calculated in section \ref{sec:spectrum}.
We discuss our results in the context of past and future LIGO searches
in section \ref{sec:discussion}.

\section{Mass quadrupole moment of a magnetic mountain}
\label{sec:mountains}
\begin{figure*}
  \includegraphics[width=168mm, keepaspectratio]{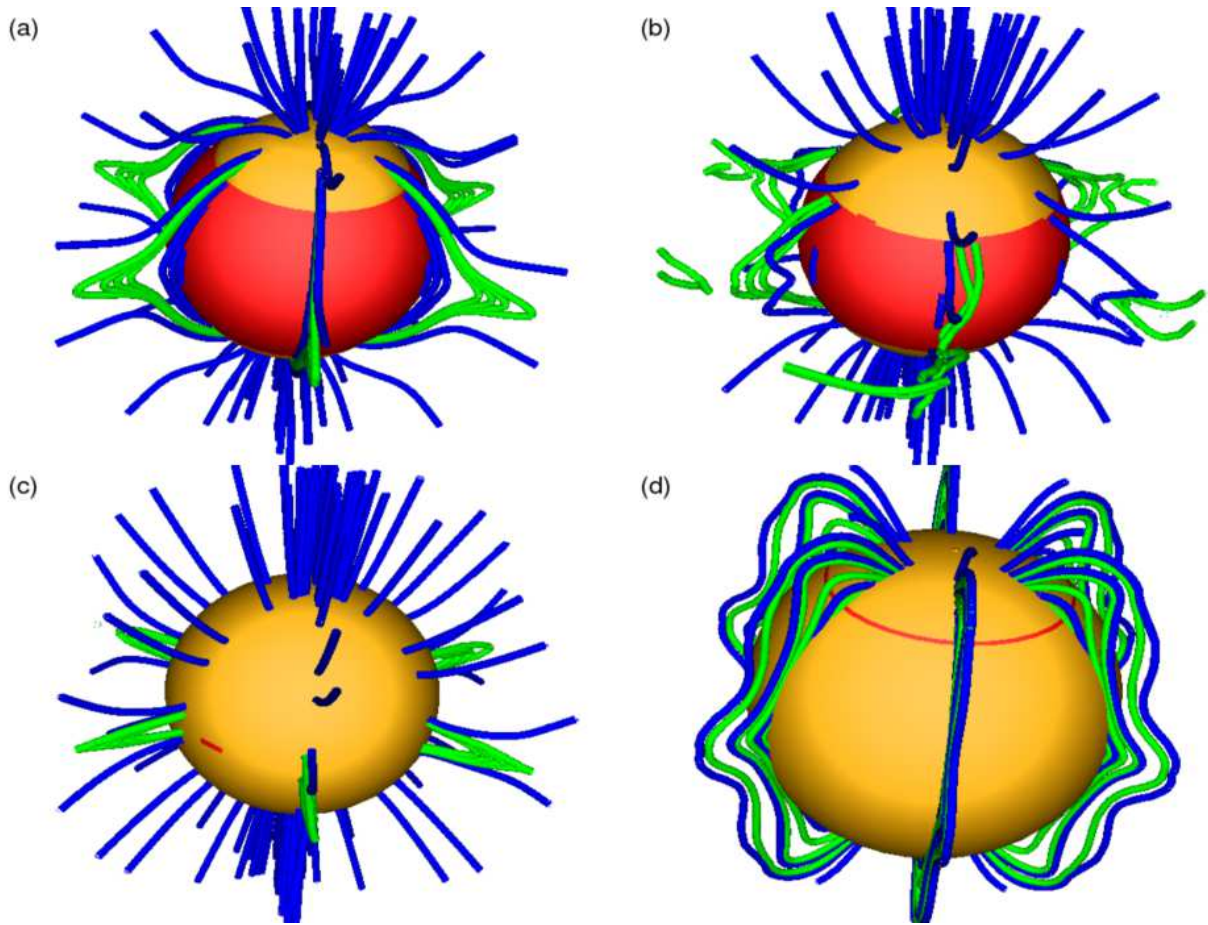}
  \caption{Density isosurface (orange) and magnetic field lines (blue
    and green) for a magnetic mountain. (a) Axisymmetric, ideal-MHD
    configuration with $M_a=1.2\times10^{-4} M_\odot$. (b)
    Nonaxisymmetric, ideal-MHD configuration with $M_a=1.2\times10^{-4}
    M_\odot$. (c) Snapshot of the resistive evolution at $t=10 \tau_\mathrm{A}$,
    with $M_a = 1.2 \times 10^{-4} M_\odot$. (d) A mountain grown by
    injection with $M_a=1.9 \times 10^{-3} M_\odot$. The
    mountain is  defined by the orange isosurface
    $\rho(r,\theta,\phi)=0.5 \rho_\mathrm{max}$, where
    $\rho_\mathrm{max}=2.0 \times 10^9$ g cm$^{-3}$ is reached at
    $\tilde{x}=(r-R_\ast)/h_0=0.9 \times 10^{-3}$ and $\theta=0.01$ in
    the axisymmetric model. In order to
    aid the reader, the altitude scales are magnified
    five-fold in all panels. The foot points of the blue field lines touch the 
    stellar surface, while green field lines are traced starting from
    the equator.}  
  \label{fig:mountains}
\end{figure*}

In the context of gravitational radiation, the key property of a
magnetic mountain is its mass quadrupole moment,
\begin{equation}
  Q_{ij}=\int d^3 \mathbf{x}' \, (3 x_i' x_j'-r'^2 \delta_{ij}) \rho(\mathbf{x'}),
\end{equation}
where $\rho$ denotes the plasma density. We aim to calculate $Q_{ij}$
as a function of the accreted mass $M_a$. It is useful to measure
$M_a$ in units of the critical mass $M_c=G M_\ast B_\ast^2 R_\ast^2/(8
c_s^4)$, where $M_\ast$ and $R_\ast$ are the stellar mass and radius,
$B_\ast=10^{12}$ G is the initial magnetic field, and $c_s=10^8$ cm
s$^{-1}$ is the isothermal sound speed. For $M_a>M_c$, the
magnetic dipole moment $\mu$ decreases with $M_a$; for $M_a<M_c$,
$\mu$ is approximately constant \citep{Payne04}. In a typical LMXB, we
have $10^{-5} \la M_c/M_\odot \la 10^{-4}$ and $10^{-2} \la M_a/M_\odot \la
10^{-1}$; that is, magnetic burial distorts the field dramatically.

The three-dimensional equilibria computed by \citet{Vigelius08a}
deviate from axisymmetry (with respect to the \emph{magnetic} axis) by less than 0.1 per cent
in the mass quadrupole moment. Hence, we frequently employ the
axisymmetric mass ellipticity, defined as
\begin{equation}
  \label{eq:app:gw:ellipticity}
  \epsilon=\frac{\pi}{I_{\hat{z}\hat{z}}}\int d\theta\,dr\, \rho r^4 \sin \theta  (3 \cos^2
  \theta - 1),
\end{equation}
to describe the mountain as a biaxial ellipsoid, where
$I_{\hat{z}\hat{z}}=2 M_\ast R_\ast^2/5$ denotes the unperturbed
moment of inertia.

\change{In order to give a reliable estimate of the magnitude of
  $\epsilon$ and hence the strength of the gravitatational wave
  signal, we take into account the effects of hydromagnetic \citep{Vigelius08a} and
  resistive \citep{Vigelius08b} relaxation and consider the high-$M_a$
  limit as well as the influence of the neutron star curvature. For
  the convenience of the reader, we repeat previously published
  results in the first three subsections.}
We compute $Q_{ij}$ for the three-dimensional equilibrium state in
section \ref{sec:mountains:equilibrium}. We then examine how the
mountain responds to hydromagnetic and resistive relaxation in sections
\ref{sec:mountains:hydro} and \ref{sec:mountains:resistive}
respectively. Section \ref{sec:mountains:late_stage} explains how to
build mountains with realistic values of $M_a$, and the effect of curvature
downscaling is quantified in section
\ref{sec:mountains:curvature}. \change{The results in the last two subsections
are new and have not been published elsewhere.}

\subsection{Three-dimensional equilibrium}
\label{sec:mountains:equilibrium}
In a magnetic mountain at equilibrium, the pressure gradient balances
the gravitational and Lorentz forces. Starting with a centred
magnetic dipole before accretion begins, \citet{Payne04}
computed the unique, self-consistent, Grad-Shafranov equilibrium that
satisfies force balance, while simultaneously respecting the
flux-freezing constraint of ideal magnetohydrodynamics (MHD).

The top-left panel of \fref{mountains} displays an axisymmetric
equilibrium for $M_a = M_c$. The mountain (orange isosurface) is
confined to the magnetic pole by the tension of the distorted magnetic
field (blue and green curves). Blue and green field lines are drawn
starting from the pole and equator, respectively. The region where the
magnetic pressure is greatest (at $\theta \approx \pi/4$ in
Fig. \ref{fig:mountains}a) is termed the equatorial magnetic
belt. Here, $B$ is $\sim 16$ times higher than at the pole.

\begin{figure}
  \includegraphics[width=84mm, keepaspectratio]{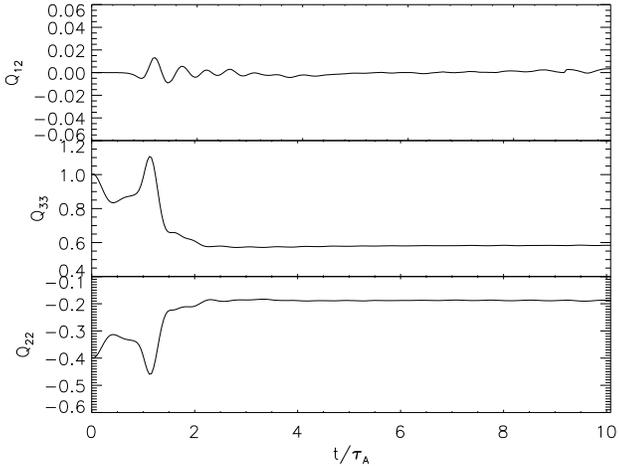}
  \caption{Quadrupole moments for the
    nonaxisymmetric configuration (Fig. \ref{fig:mountains}b), normalised to $Q_{33} =
    1.30\times10^{25}$ g cm$^2$ of the axisymmetric model, as a
    function of time, in units of the
    Alfv\'{e}n time, $\tau_\mathrm{A}=5.1\times10^{-2}$ s. The
    system develops an appreciable nonaxisymmetry,
    characterised by the off-diagonal element $Q_{12}$, during the
    relaxation phase before settling into a
    nearly axisymmetric state.}
  \label{fig:modii_quadru}
\end{figure}

However, an axisymmetric analysis neglects important toroidal
modes. When we load the axisymmetric equilibrium in
Fig. \ref{fig:mountains}a into the ideal-MHD code \textsc{zeus-mp},
we observe that it is unstable to the undulating submode of the
three-dimensional Parker instability, which reconfigures the
hydromagnetic structure by growing the toroidal magnetic field
\citep{Vigelius08a}. Thanks to line tying, the instability is not
disruptive; its saturation state (\fref{mountains}b) still confines
matter at the magnetic pole, and the mountain is still present
even though $\epsilon$ is reduced to $\sim 60$ per cent of its
original value. The evolution of $Q_{ij}$ for this model
(\fref{mountains}b) is displayed in \fref{modii_quadru}. The diagonal
components, describing the axisymmetric distortions, decrease by $\approx$ 40 per cent,
on a time-scale of $\approx 2 \tau_\mathrm{A}$. Here,
$\tau_\mathrm{A}=(\pi R_\ast \rho^{1/2}/B)_\mathrm{min}$ denotes
the pole-equator crossing time for transverse Alfv\'{e}n waves; $\pi
R_\ast \rho^{1/2}/B$ is smallest close to the stellar surface. For a
realistic star with $M_a=M_c$, the
Alfv\'{e}n time-scale evaluates to $\tau_\mathrm{A}=5.1 
\times 10^{-2}$ s.

\subsection{Fast, hydromagnetic relaxation}
\label{sec:mountains:hydro}
A magnetic mountain performs global hydromagnetic oscillations when
perturbed, but it remains intact. This unexpected outcome can be
ascribed to two factors: (i) the mountain is already the saturation
state of the nonlinear Parker instability, and (ii) the line-tying
at the stellar surface suppresses important localised modes,
e.g. interchange modes. \citet{Payne07} found all mountains with $M_a
\le 6\times 10^{-4} M_\odot$ to be marginally stable.

The mountain quickly tends to an almost axisymmetric state
($|Q_{12}/Q_{33}|<10^{-3}$ in \fref{modii_quadru}). This high degree of
axisymmetry considerably simplifies the computation of the amplitude
of the gravitational wave strain (see \sref{gravwaves:emission}). (Note
that $Q_{ij}$ is defined relative to the magnetic axis, which is
inclined with respect to the rotation axis). The reconfiguration
is accompanied by global, nonaxisymmetric, MHD oscillations (top panel
of \fref{modii_quadru}). Although they die 
away in this numerical experiment, global oscillations can be continuously
excited in reality (e.g. by accretion
torques),  modifying the gravitational wave spectrum (see \sref{spectrum}).
\citet{Payne07} identified two dominant modes: a
short-period sound mode, with a frequency
$f_\mathrm{S}/\mathrm{kHz}=1.4 \times 10^5 (c_s/10^8 \mathrm{cm\,s}^{-1})$
(independent of $M_a$), and a longer period Alfv\'{e}n oscillation,
which can be fitted by $f_\mathrm{A}=17 (M_a/M_c)$ Hz.

\begin{figure*}
  \begin{center}
    \includegraphics[width= \textwidth, keepaspectratio]{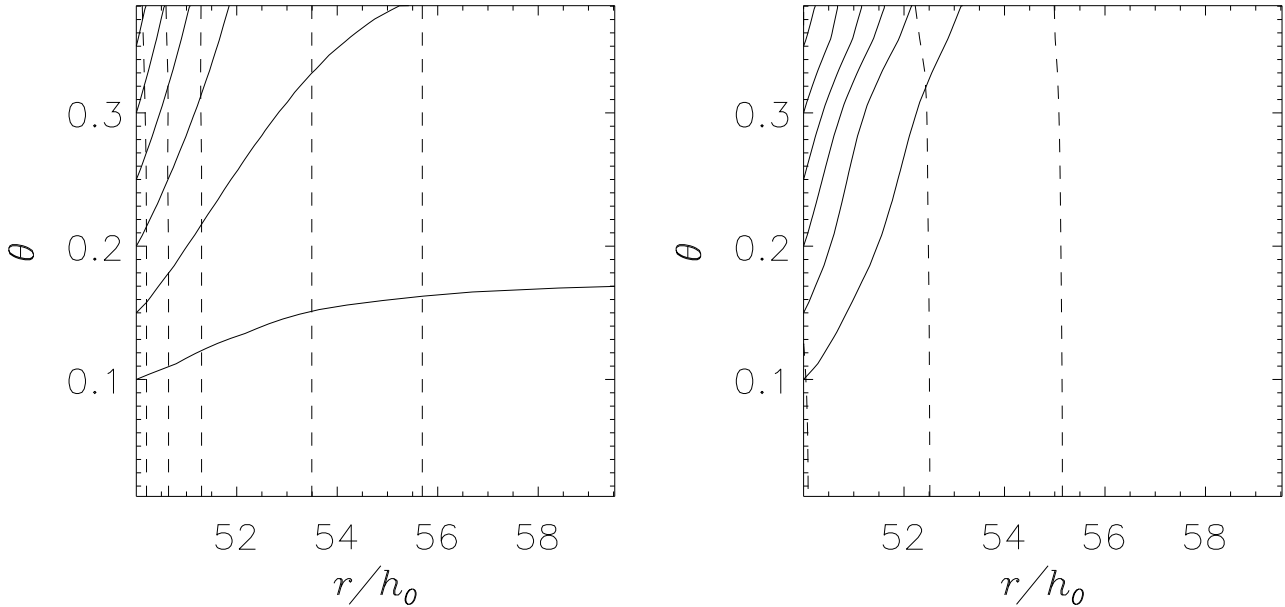}
  \end{center}
  \caption{\change{Meridional section of a mountain with $M_a=M_c$ and \texttt{outflow}
  boundary conditions at $\theta=\pi/8$ at $t=0$ (left panel) and $t=0.9
  \tau_\mathrm{A}$ (right panel). The mountain is susceptible to an
  ideal-MHD ballooning instability and disrupts over the Alfv\'{e}n
  timescale.}} 
  \label{fig:ballooning}
\end{figure*}

\change{\citep{Litwin01} demonstrated that an accretion column that is
  magnetically confined to the polar cap is susceptible to the
  ideal-MHD ballooning instability. However, our simulations do not exhibit any evidence
for a growing instability. Physically, this is because the compressed equatorial magnetic
field stabilises the lateral motions involved in such an
instability. Indeed, in \fref{ballooning}, we demonstrate
that a magnetically confined mountain is only 
susceptible to a growing ballooning mode when the back-reaction of the
magnetic belt is neglected. We perform an axisymmetric
simulation with $M_a=M_c$ and $0 \le \theta \le \pi/8$, where the
outer $\theta$-boundary (at $\theta=\pi/8$) is set to \texttt{outflow}. The mountain
is clearly disrupted on the Alfv\'{e}n timescale by the ballooning
mode, as in \fref{ballooning}; the magnetic field and
frozen-in plasma slide sideways through the outflow boundary. This is
consistent with the findings of \citet{Litwin01}, who imposed boundary
conditions equivalent to \textrm{outflow}, therefore neglecting the
stabilizing effect of the equatorial magnetic belt. Furthermore, our
growing simulations do not show any evidence for an instability during
the early stages of accretion (when $M_a \ll M_c$).}

\subsection{Slow, resistive relaxation}
\label{sec:mountains:resistive}
\begin{figure*}
  \includegraphics[width=168mm, keepaspectratio]{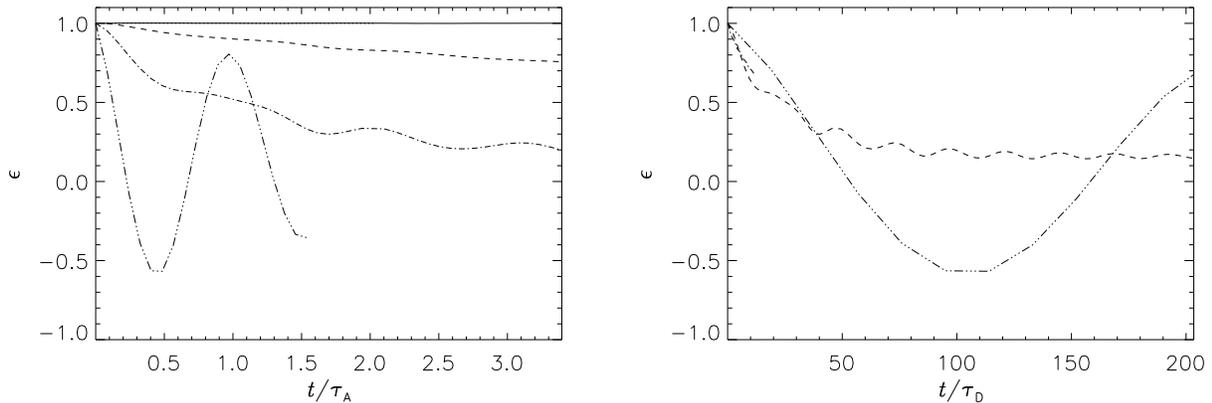}
  \caption{Evolution of mass ellipticity $\epsilon$ for different
    values of the Lundquist number
    $Lu=\tau_\mathrm{D}/\tau_\mathrm{A}=10^{14}, 10^3, 10^{-1}, 10^{-2}, 10^{-3}$ (solid, dotted, dashed,
    dash-dotted, dash-triple-dotted), curves from top to bottom, in units of the Alfv\'{e}n
    time $\tau_\mathrm{A}=2.5 \times 10^{-21}$ s (left panel) and the
    respective diffusion times $\tau_\mathrm{D}$ (right panel). The
    solid and dotted curves nearly overlap. The
    magnetic mountain relaxes at the time-scale $\tau_\mathrm{D}$,
    with $\epsilon$ falling to $\mathrm{e}^{-1}$ of its initial value after
    $34 \tau_\mathrm{D}$. Two of the curves are hard to see in the
    right panel: $Lu=10^{14}$ (solid) stops at $t=\tau_\mathrm{D}$ and
    $Lu=10^3$ (dotted) stops at $t=10 \tau_\mathrm{D}$.}
  \label{fig:ell_with_time_3d}
\end{figure*}

A magnetic mountain relaxes resistively over a long time-scale, which
is set by
the ohmic diffusion rate across the steepest magnetic gradients in
the mountain. Resistive relaxation reduces $Q_{ij}$. Simple estimates
suggest that the effect can be neglected as long as $M_a \la 10^{-5}
M_\odot$, assuming a homogeneous conductivity,  $\sigma=7.7 \times
10^{26}$ s$^{-1}$ \citep{Melatos05}. Here, we assume that 
$\sigma$ is dominated by electron-phonon scattering, with a crustal
temperature $T=10^7$ K and a characteristic plasma density of
\cchange{$\rho=5\times 10^{13}$ g cm$^{-3}$} \citep{Cumming04}. \cchange{There
  are considerable uncertainties about the exact value of $\sigma$,
  such as the value of the impurity parameter \citep{Schatz99,
    Cumming01, Cumming04, Jones04, Pons07} and the composition of the
  crust \citep{Cumming04, Chamel08}. Furthermore, $\sigma$ is in reality
  a function of the location through $T$ and $\rho$. In
  keeping with \citet{Vigelius08b}, we treat the electrical
  conductivity as a fiducial parameter and note that the time-scale of
  resistive relaxation scales with $\sigma$. } An
inhomogeneous conductivity will be considered in a forthcoming paper.

Transient resistive instabilities, like a global tearing mode or
local gravitational mode \citep{Furth63}, are known to evolve on
time-scales as short as $\sim (\tau_\mathrm{D}
\tau_\mathrm{A})^{1/2} \approx 0.3\; \mathrm{yr}$, where
$\tau_\mathrm{D}$ denotes the characteristic
diffusion time-scale.  \change{Note that $\tau_\mathrm{D}$ implicitly
  depends on the position through $\bmath{B}$ and $\rho$ and has a
  minimum close to the surface. The definition of the characteristic
  diffusion timescale is hence somewhat arbitrary. As justified in
  \sref{mountains:equilibrium}, we pick the minimum value
  $\tau_\mathrm{D}=\left(4 \pi \sigma B/c^2 |\nabla^2 
  \mathbf{B}|\right)_\mathrm{min}$.} Instabilities grow in
magnetic neutral sheets [created by the undulating submode of the
Parker instability; see \citet{Hanasz02}] or regions of high magnetic
shear. 

\citet{Vigelius08b} tested numerically whether neutral sheets can grow to
disrupt the mountain on short time-scales. They evolved the
three-dimensional equilibrium in Fig. \ref{fig:mountains}b, and
similar states for other values of $M_a$, in \textsc{zeus-mp},
extended to treat ohmic diffusion.
The results are reported in \fref{ell_with_time_3d},
which depicts the evolution of $\epsilon$ for different Lundquist numbers
$Lu=\tau_\mathrm{D}/\tau_\mathrm{A}$. For each value of $Lu$,
$\epsilon(t)$ is plotted as a function of time, measured in units of the Alfv\'{e}n
time\footnote{In \fref{ell_with_time_3d}, $\tau_\mathrm{A}$ is the
  characteristic Alfv\'{e}n time-scale for the three-dimensional equilibrium.
  It is half the characteristic Alfv\'{e}n time-scale of
  the axisymmetric configuration used in section
  \ref{sec:mountains:equilibrium}.} $\tau_\mathrm{A}=2.5 \times
10^{-2}$ s (left panel) and the diffusion time $\tau_\mathrm{D}$ (right panel), which differs for
each model. The models with a realistic resistivity ($Lu=10^{14}$,
solid curve) and $Lu=10^3$ (dotted) do not exhibit any change in $\epsilon$ over 
the simulation time. For $Lu=10^5$ (dashed), we note a decrease of 21 per cent
over $\sim 0.1 \tau_\mathrm{D}$. For $Lu=10^{-2}$ (dash-dotted), the
mountain relaxes substantially; $\epsilon$ drops by 90 per cent over the diffusion
time-scale. For $Lu=10^{-3}$ (dash-triple-dotted), the mountain immediately slips through the
magnetic field lines and falls freely towards the equator, where it is
reflected by the boundary surface, causing $\epsilon$ to
oscillate. The latter case, in particular, is of academic interest
only, as far as its application in LMXBs is concerned. Magnetic
neutral sheets are found in the toroidal plane, where
the plasma density and magnetic field strength are low. Reconnection
occurs locally in these regions, smoothing toroidal gradients. Ohmic
dissipation therefore tends to restore axisymmetry.

Importantly, the mountain relaxes globally on the diffusion time-scale,
$\tau_\mathrm{D}$, which greatly exceeds the accretion time-scale $\tau_\mathrm{acc}$,
with $\epsilon$ falling to $\mathrm{e}^{-1}$ of its initial value after $34
\tau_\mathrm{D}$. In practice,
  this means that 
the three-dimensional saturation state of the Parker instability does
not relax resistively until $\sim 10^7$ yr elapse, at least for the
examined mountains with $M_a \la 10^{-4} M_\odot$. In the analytic
small-$M_a$ limit, \citet{Melatos05} found the minimum accreted mass for
ohmic diffusion to arrest mountain growth to be $M_\mathrm{d} \sim 10^{-7}
M_\odot$, provided the star accretes at the Eddington rate (such that
$\tau_\mathrm{acc} = M_\mathrm{d}/\dot{M_a} \sim 10^5$ yr) and has a crustal
temperature of $T=10^8$ K. Our numerical result, independently
validated in \sref{mountains:late_stage}, raises $M_d$
by three orders of magnitude.

In addition to resistive effects in the accreted plasma,
\citet{Konar97} explored the influence of ohmic dissipation on the
crustal magnetic field. They identified three competing mechanisms:
(i) the accretion flow advects current
into lower layers with higher $\rho$ and consequently higher
$\sigma$; (ii) the crust is heated by advection, decreasing $\sigma$;
and (iii) the current is squeezed into the inner layers, decreasing
$\tau_\mathrm{D}$. As a result, the crustal field decays rapidly
before freezing at a residual surface value, whose magnitude increases
with $\dot{M}_a$ [by reducing the duration of the rapid decay
phase, see also \citet{Romani90}]. We defer the study of sinking and
stratified $\sigma$ \change{[compare, e.g.,
  \citep{JahanMiri00,Choudhuri02}]} to future work.

\subsection{Realistic accreted mass}
\label{sec:mountains:late_stage}
Numerical obstacles, like steep gradients, and physical obstacles,
like magnetic bubble formation, interfere with the task of modelling
magnetic mountains for realistic values of $M_a$.
The iterative numerical scheme employed by \citet{Payne04} to compute
Grad-Shafranov equilibria converges poorly for $M_a \ga 10^{-4}
M_\odot$. A bootstrapping algorithm can be used to quasistatically fatten a
Grad-Shafranov equilibrium ten-fold \citep{Payne07}, but the results
have not yet been verified against a reliable numerical solution of
the full initial-value problem (bootstrapping converges quite
violently), and anyway, bootstrapping works up to $M_a \sim 10^{-3}
M_\odot$ at most. In short, a self-consistent configuration with a
realistic amount of accreted matter, e.g. $M_a \sim 0.1 M_\odot$
\citep{Burderi99}, is yet to be achieved.

We try to overcome this restriction in this
subsection by growing a magnetic mountain from scratch, by injecting
plasma at the inner boundary $r=R_\ast$  into an initially dipolar
background field. The injection speed is chosen to be less than the
gravitational escape speed from the simulation box, while the density
is chosen to give $\tau_\mathrm{A} \ll \tau_\mathrm{acc}$, such that the
system passes through a sequence of \emph{quasistatic} equilibria,
yet $\tau_\mathrm{acc}$ is short enough to keep the simulation runtime
reasonable for $M_a \sim 10^{-3} M_\odot$. This approach differs from the bootstrapping algorithm
\citep{Payne07} in two ways: (i) instead of relying on a
Grad-Shafranov equilibrium as the starting point, we solve the full
initial-value problem; and (ii) we inject plasma at the $r=R_\ast$
boundary, thereby circumventing the artificial field line pinning at the
outer boundary that stems from the \texttt{inflow} boundary
condition. Implementation details are provided in
Appendix \ref{sec:app:grow}. 

An axisymmetric grown mountain with $M_a = 1.9 \times 10^{-3} M_\odot$
and $b=3$ (hemispheric-polar magnetic flux ratio; see Appendix
\ref{sec:app:grow}) is displayed in panel (d) of \fref{mountains}. The mountain
isosurface covers the whole star. The base density, at
$(\tilde{x},\theta)=(10^{-3},0.012)$ is fifty times higher than for
$M_a=1.2 \times 10^{-4} M_\odot$ [panel (a)]. At first glance, the magnetic field configuration looks entirely
different: instead of pointing radially outward, all field lines are
closed loops. This ostensible difference is due to the boundary
condition $\partial \mathbf{B}/\partial r=0$ at $r=R_m$, enforced in
the growing simulations (see Appendix \ref{sec:app:grow}),
cf. $B_\theta=0$ in \citet{Payne04}. However, appearances are a bit
misleading: the all-important equatorial belt, where the
magnetic field is highly distorted and most intense, is clearly
visible in \fref{mountains}d, just as much as
Figs. \ref{fig:mountains}a -- \ref{fig:mountains}c. Near the pole, at
$(\tilde{x},\theta)=(10^{-3},0.012)$, $B$ in model d is comparable to
$B$ in model a. At the magnetic equator, $(\tilde{x}, \theta)=(10^{-3},
1.4)$, $B$ is $\sim 25$ times higher in model d than in model a.
Naturally, the magnetic tension required to counterbalance the
hydrostatic pressure is greater in model d.

\begin{figure}
  \includegraphics[width=84mm,keepaspectratio]{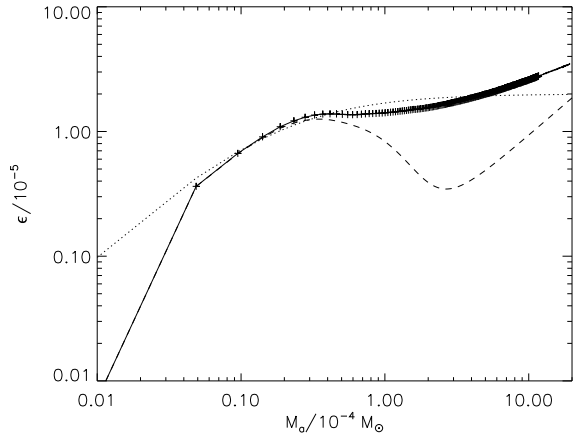}
  \caption{Mass ellipticity $\epsilon$ as a function of $M_a$. The
    solid curves show a model with zero resistivity (solid curve), realistic
    resistivity $\eta=1.3 \times 10^{-27}$ s (plus symbols),
    and an artificially high resistivity $\eta=9.2 \times 10^{-13}$ s
    (dashed curve). The dotted curve represents a fit to the (slightly
    modified) formula given in \citet{Melatos05}; see
    \eref{mountains:late_stage:fit}.}
  \label{fig:mountain:growth_ellipticity}
\end{figure}

\fref{mountain:growth_ellipticity} shows the evolution of $\epsilon$
as a function of $M_a$ up to $M_a \le 1.9\times10^{-3} M_\odot$. Plotted are
models with zero resistivity (solid curve), a realistic resistivity
$\eta=1.3 \times 10^{-27}$ s [\citet{Cumming04}, plus symbols],
and an artificially high resistivity, $\eta=9.2 \times 10^{-13}$
s (i.e. $Lu=\tau_\mathrm{D}/\tau_\mathrm{A}=10^5$, dashed
curve). We also fit a functional dependence similar to that proposed
by \citet{Shibazaki89} and \citet{Melatos05} to the plus symbols, finding
\begin{equation}
  \label{eq:mountains:late_stage:fit}
  \frac{\epsilon}{10^{-5}} = 11 \frac{M_a}{10^{-4} M_\odot} \left( 1+5.5
    \frac{M_a}{10^{-4} M_\odot} \right)^{-1}.
\end{equation}
The fitting formula is plotted as a dotted curve in
\fref{mountain:growth_ellipticity}. 

\fref{mountain:growth_ellipticity} and \eref{mountains:late_stage:fit}
are indispensable tools for calculating the gravitational wave
emission, summarising as they do our most up-to-date modelling of
resistive relaxation and large (i.e. realistic) values of $M_a$. For
$M_a \ga 0.5 \times 10^{-4} M_\odot$, $\epsilon$ is expected and found
to follow the shape of the small-$M_a$ analytic solution, valid for
$M_a \ll M_c=1.2 \times 10^{-4}$ [see Appendix in \citet{Payne04}]. The
deviation visible at $M_a=10^{-6} M_\odot$ can be attributed to the fact
that the configuration has not yet had time to equilibrate. In
this regime, material piles up in a polar flux tube of approximately
constant cross-sectional area, yielding $\epsilon \propto M_a$. For
$M_a \ga 10^{-5} M_\odot$, the hydrostatic pressure overcomes the magnetic
tension and the mountain spreads towards the equator, triggering
global MHD oscillations. These lateral oscillations, observed numerically by
\citet{Payne07}, compress
the magnetic field once per cycle, increasing $|\nabla^2 \mathbf{B}|$,
promoting lateral diffusion of the mountain, and reducing
$\epsilon$. For high $\eta$ (dashed curve), the oscillation in $\epsilon$
is clearly visible in \fref{mountain:growth_ellipticity}. It is also
visible, albeit less prominently, for realistic $\eta$ (plus symbols
in \fref{mountain:growth_ellipticity}. Hence the fairest way to
interpret the $\epsilon$-$M_a$ trend in
\fref{mountain:growth_ellipticity} is that $\epsilon$ saturates for
$M_a\gg M_c$ (flat underlying trend) with an oscillation superposed. 

In a realistic accretion scenario, given
$\dot{M_a}$, we expect to find one particular value of $M_a$ (and hence
$\epsilon$) at which the system attains a steady state, where
the mass diffuses through flux surfaces at a rate which is exactly
replenished by $\dot{M_a}$. A full parameter study to examine the
relation $\epsilon(\dot{M}_a)$ is outside the scope of this article. We simply note,
following \citet{Vigelius08b}, that magnetic mountains are resistively
stable over $\tau_\mathrm{acc}$ or $M_a \le 1.2 \times 10^{-4}
M_\odot$ (neglecting mass inflow).

\change{Can short-timescale instabilities that are absent in the
  equilibrium configuration grow during the early stages of accretion,
  i.e. for small $M_a$? In 
  this case, high-$M_a$ equilibria could never be reached and magnetic
  mountains would not emit detectable gravitational
  radiation. However, our growing simulations
  (\fref{mountain:growth_ellipticity}) show no evidence for 
  ideal-MHD or resistive instabilities during the low-$M_a$ stage of
  accretion.}

\subsection{Realistic curvature}
\label{sec:mountains:curvature}
In general, the characteristic length-scale for radial gradients
($h_0$) is much smaller than the length-scale for
latitudinal gradients ($R_\ast$), creating numerical difficulties. However, in
the small-$M_a$ limit, it can be shown analytically \citep{Payne04, Payne06a} that
the structure of the magnetic mountain depends on
$R_\ast$ and $M_\ast$ through the combination $h_0 \propto
R_\ast^2/M_\ast$, not separately. We therefore artificially reduce $R_\ast$ and $M_\ast$, while
keeping $h_0$ fixed, to render the problem tractable computationally. It is
vital to bear in mind that invariance of the equilibrium
structure under this curvature rescaling does not imply invariance
of the dynamical behaviour, nor is the scaling necessarily applicable at large $M_a$.

A standard neutron star has $M_\ast=1.4 M_\odot$,
$R_\ast=10^6$ cm, $B_\ast=10^{12}$ G, and $c_s=10^8$ cm s$^{-1}$. We
rescale the star to $M_\ast'=1.0\times10^{-5} M_\odot$ and
$R_\ast'=2.7 \times 10^3$ cm, reducing $a=R_\ast/h_0$ to 50 while keeping it
large. We then calculate the mountain structure, and hence $\epsilon$
numerically. We upscale $M_a$ back to a realistic star, using the
scaling relation for $M_c$, the critical accreted mass above which the
star's magnetic moment starts to change, defined by equation
(30) of \citet{Payne04}:
\begin{equation}
  \frac{M_c}{M_\odot} = 6.2 \times 10^{-15} \left(\frac{a}{50}\right)^4
 \left( \frac{B_\ast}{10^{12} \mathrm{G}} \right)^2   
  \left( \frac{c_s}{10^8 \mathrm{cm\;s}^{-1}} \right)^{-4}.
\end{equation}
Furthermore, we use the analytic result $\epsilon \propto a^2$
\citep{Melatos05}, valid for $M_a \ll M_c$, to upscale $\epsilon$. In
order to verify the fairness of this procedure, we
perform runs for $a=100$ and $a=500$. If the curvature rescaling is
fair, these runs should obey $\epsilon \propto a^2$. Computational costs
limit us to achieving $M_a = 0.34\times10^{-4} M_\odot$. We
compute $\epsilon$, upscaled to a realistic star, for the maximum $M_a$ for
both runs, finding a relative deviation between the simulation output
and the predicted scaling of $<10^{-5}$ per cent.

\section{Gravitational wave strain}
\label{sec:gravwaves}

\subsection{Ellipticity}
\label{sec:gravwaves:emission}
The neutron star and the piled up matter at the magnetic pole
can be modelled approximately as a rigid, biaxial top, which is
symmetric about the pre-accretion magnetic axis. Biaxial equilibria
are of course unstable to the toroidal Parker mode and reconfigure into a triaxial
equilibrium, as described in \sref{mountains:hydro} and
\citet{Vigelius08a}. However, the ultimate deviations from axisymmetry are
small, less than $0.1$ per cent, according to \fref{modii_quadru}, and
can be neglected in a first analysis. In this case, we can rewrite
\eref{app:gw:ellipticity} as  $\epsilon=|I_3-I_1|/I_1$,
where $I_1$ and $I_3$ denote the moments of inertia with respect to the star's
principal axes. It is important to keep in mind that we are
dealing with a \emph{prolate} spheroid, with $I_1>I_3$; the
consequences for the long-term rotational evolution are explored in
\sref{spectrum:triaxial}. Frequently, authors omit taking the absolute value in the
definition of $\epsilon$ resulting in a negative ellipticity for our
case.

In general, the mountain axis is tilted with respect to the rotation
axis. Thus the neutron star precesses freely, generating gravitational
waves at $f_\ast$ and $2 f_\ast$, where $f_\ast=J/2 \pi I_3$ is the
star's spin frequency. For a biaxial star, the wave strains in orthogonal ($+$ and
$\times$) polarisations can be written as \citep{ZSI, JKSI}
\begin{eqnarray}
  \label{eq:emission:signal}
  h_+(t) & = & \frac{1}{8} h_0 \sin 2\theta \sin 2 i \cos \Phi(t) \\
 & & + \frac{1}{2} h_0 \sin^2 \theta (1+\cos^2 i) \cos 2
  \Phi(t), \nonumber \\ 
  \label{eq:emission:signal2}
  h_\times(t) & = & \frac{1}{4} h_0 \sin 2\theta \sin i \sin
    \Phi(t)  \\ 
    & & + h_0 \sin^2 \theta \cos i \sin 2 \Phi(t). \nonumber
\end{eqnarray}
Here, $\theta$ denotes the wobble angle (between the total angular
momentum $\mathbf{J}$ and principal axis of inertia $\mathbf{e}_3$), $i$
is the inclination angle (between $\mathbf{J}$ and the line of sight,
drawn from the star to the solar system barycenter), $h_0$ is a
characteristic amplitude,
\begin{equation}
  \label{eq:emission:h0}
  h_0 = \frac{16 \pi^2 G}{c^4} \frac{\epsilon I f_\ast^2}{D},
\end{equation}
(where $I$ is the moment of inertia and $D$ is the
distance to the source), and $\Phi(t)$ is the phase, including Doppler
terms; for a source at rest relative to the observer, we can write
$\Phi(t)=2 \pi f_\ast t+\Phi_0$.

Upon combining all the results in section \ref{sec:gravwaves},
principally \eref{mountains:late_stage:fit},
\fref{mountain:growth_ellipticity} and the multiplication $\approx
0.6$ for converting $\epsilon$ from two- to three-dimensional
equilibria (section \ref{sec:mountains:equilibrium}), we arrive at the
following approximate formula for the ellipticity in the absence of
resistivity:
\begin{equation}
  \label{eq:emission:epsilon_fit}
  \frac{\epsilon}{10^{-5}} = 6.82 \frac{M_a}{10^{-4} M_\odot} \left( 1+5.5
    \frac{M_a}{10^{-4} M_\odot} \right)^{-1}.
\end{equation}

Resistive relaxation arrests the growth of the mountain (and hence
$\epsilon$) at a value of $M_a$, denoted $M_d$, which depends on $\dot{M_a}$ (see
\sref{mountains:late_stage}). Conservatively, we conclude from the
results in \sref{mountains:resistive} that a realistic
resistivity does not relax a mountain with $M_a=1.2 \times 10^{-4}
M_\odot$ over the accretion time-scale. This implies $\epsilon \le 3.6
\times 10^{-4}$, but the true value of $\epsilon$ is expected to be
much lower than the upper bound.

\subsection{Signal-to-noise ratio}
\label{sec:gw:detectability}

The signal $x(t)$ read out at the detector port is buried in
noise. We assume here that the noise $n(t)$ is additive, stationary,
and Gaussian, with $\langle n(t) \rangle=0$. Then we can write \citep{JKSI}
\begin{equation}
  x(t)=h(t)+n(t),
\end{equation}
where
\begin{equation}
  \label{eq:emission:detector_signal}
  h(t)=F_+(t) h_+(t) + F_\times (t) h_\times(t)
\end{equation}
contains the beam pattern functions $F_+$ and $F_\times$, which encode the
diurnal motion of the Earth.

\change{By averaging over sky position,
inclination, and polarisation, the signal-to-noise ratio can be
expressed as a function of the wobble angle, $h_0$, and the
observation time $T_0$, viz.}
\begin{equation}
  \label{eq:sig:snr1_av}
  \left<d_1^2\right> = \frac{h_0^2 T_0 \sin^2 2 \theta}{100 S_h(f_\ast)},
\end{equation}
and
\begin{equation}
  \label{eq:sig:snr2_av}
  \left<d_2^2\right> =  \frac{4 h_0^2 T_0 \sin^4 \theta}{25 S_h (2 f_\ast)},
\end{equation}
assuming that the interferometer arms are perpendicular. \change{Here,
  $S_h(f)$ denotes the one-sided spectral noise density of the detector.}

It is sometimes desirable to average over wobble angle as well, in the
absence of knowledge about a specific object. Following
\citet{Payne06a}, we average in a manner that
is unbiased towards small $\theta$, viz. $\left<\ \right> =
\int_0^1 \ \; \mathrm{d}(\cos \theta)$; cf. \citet{Thorne87}. The
final result is
\begin{equation}
  \left<d^2\right> =\frac{2 h_0^2 T_0}{375} \left[
    \frac{1}{S_h(f_\ast)} + \frac{16}{S_h(2 f_\ast)} \right].
\end{equation}

\subsection{LIGO detectability}
\label{sec:gw:det}
\begin{figure*}
  \includegraphics[width=168mm,keepaspectratio]{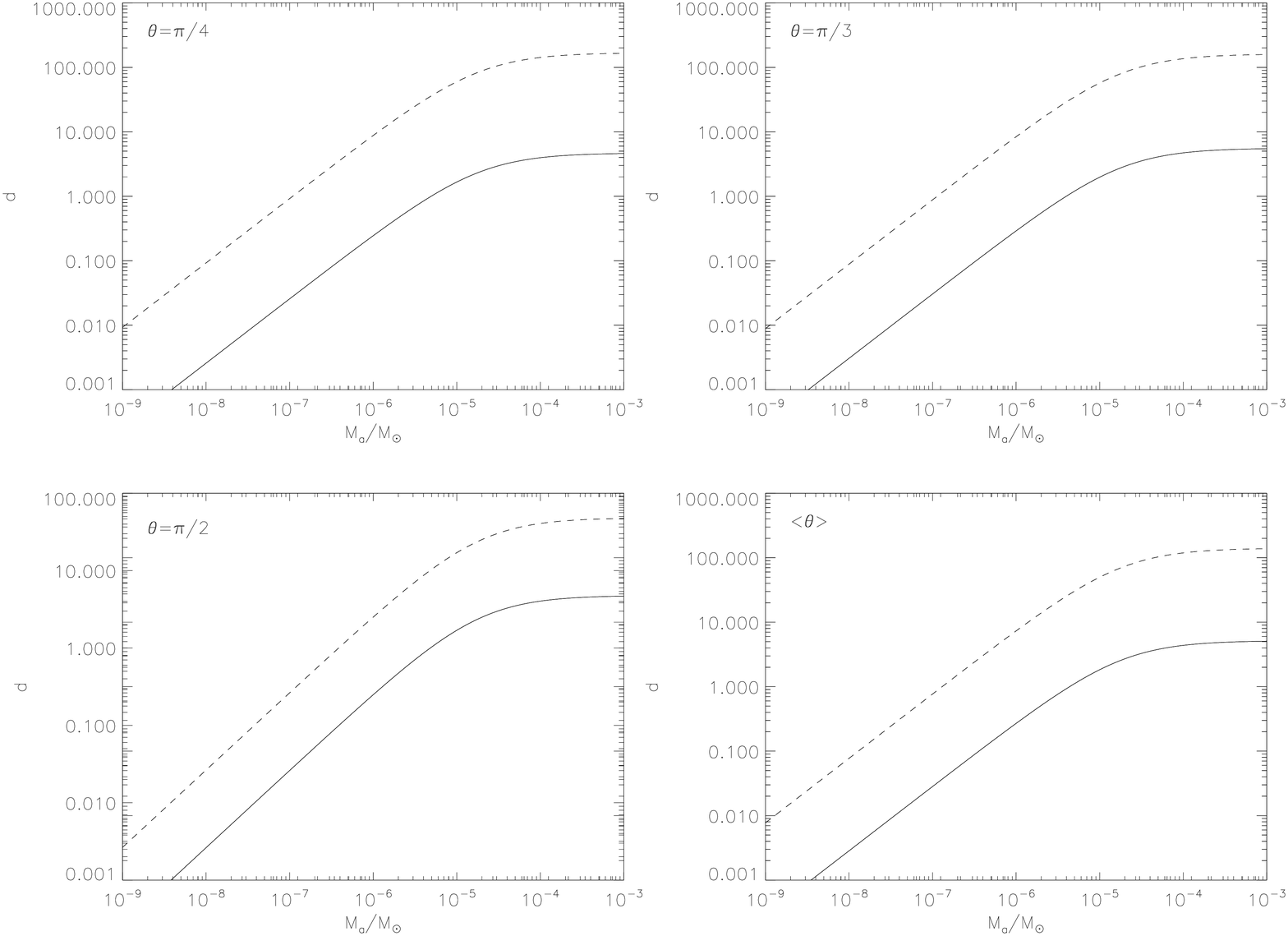}
  \caption{Signal-to-noise ratio $d$ for $10^{-9} \le M_a/M_\odot \le
    10^{-3}$ for the nonaxisymmetric equilibrium for Initial LIGO
    (solid curve) and Advanced LIGO (dashed curve) after 14 days of
    coherent integration. The wobble angles are: $\theta=\pi/4$ (top
    left), $\theta=\pi/3$ (top right), $\theta=\pi/2$ (bottom left),
    unbiased average over $0 \le \theta \le \pi/2$ (bottom right).}
  \label{fig:emission:snr}
\end{figure*}

In \fref{emission:snr}, we plot $d$ as a function of accreted mass
$M_a$ and wobble angle $\theta$ for a standard
pulsar with $f=500$ Hz, located at a distance $D=10$ kpc, assuming a
coherent integration time of $T_0=14$ d. The inputs are
\eref{emission:h0}, \eref{emission:epsilon_fit}, \eref{sig:snr1_av},
and \eref{sig:snr2_av}. We neglect resistive relaxation in order to
obtain an upper limit on $d$; the results of
\sref{mountains:resistive} and \sref{gravwaves:emission} suggest Ohmic
diffusion saturates $d$ for $M_a \ge 10^{-4} M_\odot$. The solid and
dashed curves in \fref{emission:snr} refer to the current and planned
sensitivities of Initial and Advanced LIGO respectively, published in
LIGO science requirement document and the Advanced LIGO
proposal\footnote{\texttt{http://www.ligo.caltech.edu/advLIGO/scripts/ref\_des.shtml}}. We
find that a perfectly radiating mountain with 
$M_a \ge 10^{-4} M_\odot$ is barely detectable with Initial
LIGO ($d \approx 1$) and firmly detectable with
Advanced LIGO ($d \approx 10$). As $M_a$ increases, $d$ saturates at
$\sim 10$ for LIGO and $\sim 100$ Advanced LIGO. One must remember,
though, that resistive relaxation reduces these values for $M_a \ga
10^{-4} M_\odot$.

\begin{figure}
  \includegraphics[width=84mm,keepaspectratio]{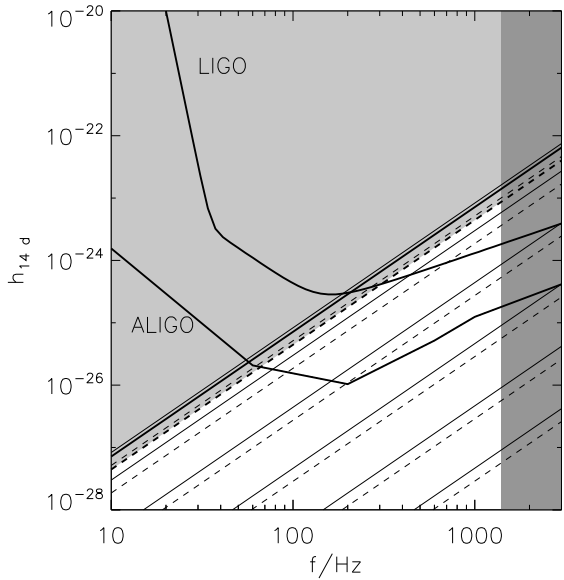}
  \caption{Amplitude $h_0$ of the gravitational wave signal for $M_a/M_\odot= 10^{-9},
     10^{-8},10^{-7},10^{-6},10^{-5},10^{-4},10^{-3}$, for
     axisymmetric (solid diagonal lines, bottom to top) and nonaxisymmetric
     (dashed lines) mountains. The sensitivities of Initial and
     Advanced LIGO, assuming 14 days coherent integration, are also
     plotted (upper and lower curves respectively).
     The growth of the mountain is arrested for $M_a
     \ga  10^{-4} M_\odot$ (light shaded region)
     by resistive relaxation \citep{Melatos05, Vigelius08a}. The
     right-hand edge is excluded at
     present because no accreting millisecond pulsars have been
     discovered with $2 f_\ast > 1.4$ kHz (dark shaded region).}
  \label{fig:emission:detectability}
\end{figure}

An alternative way to estimate the detectability of the signal is to
compare the characteristic wave strain $h_0$ versus the statistical
threshold $h_\mathrm{th}$. A signal is detected with a false alarm
rate of 1 per cent and a false dismissal rate of 10 per cent when
$h_0$ exceeds $h_\mathrm{th} \approx 11.4 [S_h(2 f_\ast)]^{1/2} T_0^{-1}$
\citep{JKSI, Abbott2004}. \fref{emission:detectability} displays
$h_0$ as a function of $M_a$ for the same pulsar as in
\fref{emission:snr}. The diagonal lines give $h_0$ for $M_a/M_\odot = 10^{-9},
10^{-8},10^{-7},10^{-6},10^{-5},10^{-4},10^{-3}$, for a
biaxial (solid lines) and a triaxial (dashed lines) mountain. Also
plotted as solid curves are $h_\mathrm{th}$ for Initial LIGO and
Advanced LIGO for $T_0=14$ d, averaged over $\theta$ and $i$. We exclude the region $M_a \ga 10^{-4}
M_\odot$ (light shaded region), in which resistive relaxation prevents
further growth of $\epsilon$, and the region $2 f > 1.4$ kHz (dark shaded
region), because no accreting millisecond pulsars have been discovered
yet in this band (they may be in the
future).

\fref{emission:detectability} suggests that there is a small
region $400 \la f_\ast/\mathrm{Hz} \la 1400$ for which magnetic
mountains are, in principle, detectable with initial LIGO. However,
there are several physical mechanisms not yet considered in our
modelling, most notably the sinking of the mountain into the crust,
which act to reduce the gravitational wave signal. We
discuss these mechanics further in \sref{discussion}.

\section{Gravitational wave spectrum}
\label{sec:spectrum}
In this section, we investigate the gravitational wave spectrum
in more detail. Two effects modify
the spectrum away from its simplest form (delta functions at $f_\ast$
and $2 f_\ast$): global MHD oscillations, and triaxiality. We study
the former without the latter in \sref{spectrum:biaxial}, and vice
versa in \sref{spectrum:triaxial}, to isolate the physics of the two
effects. 

\subsection{Biaxial, vibrating mountain}
\label{sec:spectrum:biaxial}
A magnetically confined mountain oscillates when plucked, e.g. by
starquakes or fluctuations in the accretion torque \citep{Payne06a,
  Vigelius08a}. These global hydromagnetic modes appear in the
gravitational wave spectrum as acoustic and Alfv\'{e}nic sidebands
beside the two main peaks at $f_\ast$ and $2 f_\ast$.

\begin{figure*}
  \includegraphics[width=168mm, keepaspectratio]{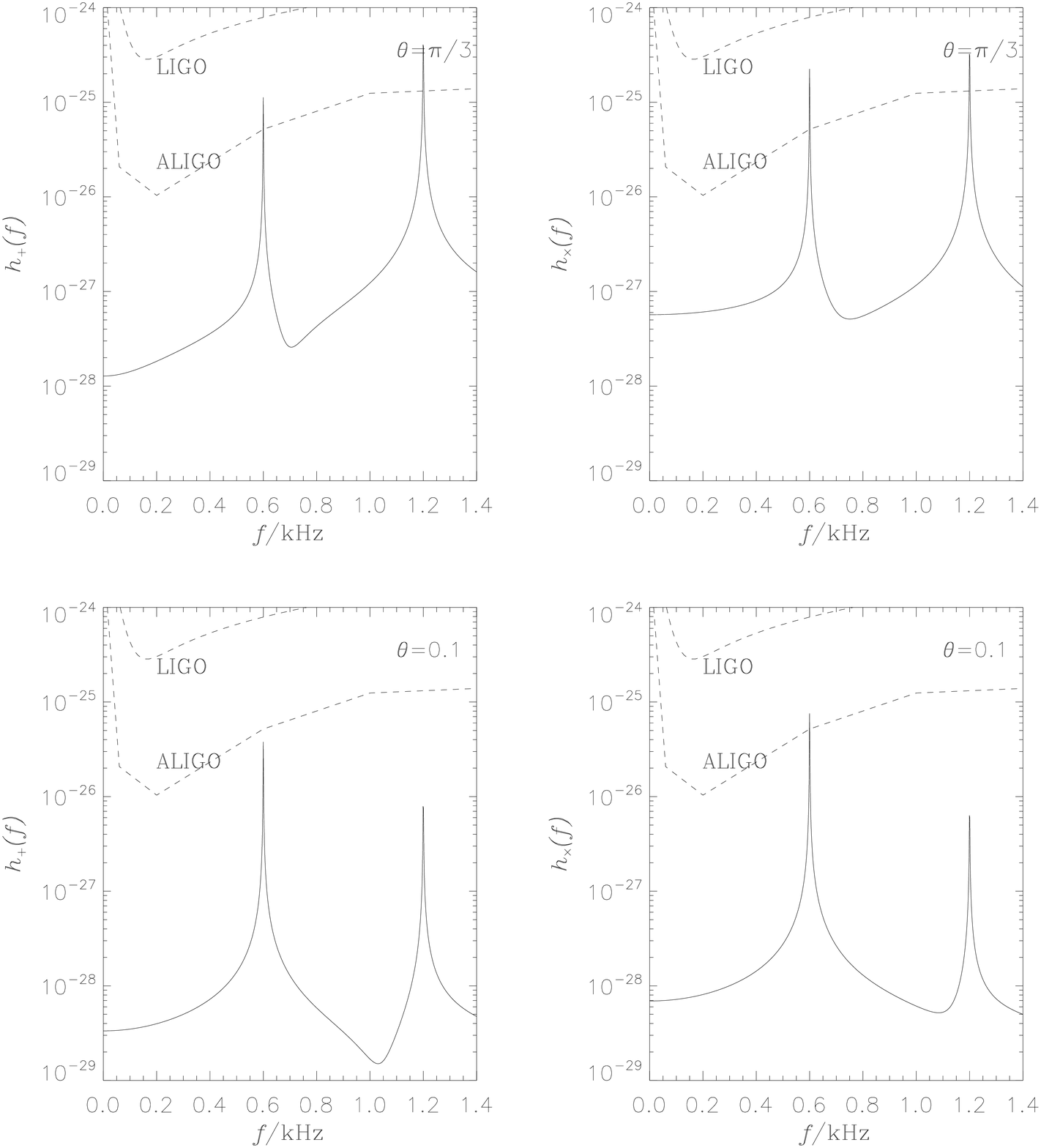}
  \caption{Fourier transform of the gravitational wave signal from a standard
    neutron star ($M_a=1.4 M_c$) with an axisymmetric magnetic mountain at a distance
    $D=10$ kpc, with wobble angle $\theta=\pi/3$
    (top panels) and $\theta=0.1$ (bottom panels),
    and inclination $i=\pi/3$. The left (right)
    panels show the discrete Fourier transforms of the signals $h_+(t)$
    [$h_\times(t)$] in the plus (cross) polarisations.
    The peak is not a $\delta$ function due to the
    finite resolution of the discrete Fourier transform. The dashed curves
    display the detection threshold $h_\mathrm{th}$ for Initial
    (upper) and Advanced (lower) LIGO, assuming an
    observation time of $T_0=14$, a false alarm rate of 1 per cent, and
    a false dismissal rate of 10 per cent.}  
  \label{fig:spectrum:axisym}
\end{figure*}

The top panels of \fref{spectrum:axisym} display the truncated Fourier transform,
\begin{equation}
  h(f) = \frac{1}{T_0}\int_{0}^{T_0}
  \mathrm{d}t\,\mathrm{e}^{i 2 \pi f t}h(t),
\end{equation}
of the wave strains  in the $+$ and $\times$
polarisations, given by equations \eeref{emission:signal} and
\eeref{emission:signal2}, for a standard star with $M_a=1.2 \times
10^{-4} 
M_\odot$ at $D=10$ kpc, with $\theta=\pi/3$ and $i=\pi/3$, assuming an integration time
of $T_0=14$ days. For comparison, we also plot $h_\mathrm{th}$.
It is important to bear in mind that, for detection, one uses the
combined signal power in both polarisations, even when the peaks in each
single polarisation remain under the threshold. The wobble angle is chosen
artificially large to illustrate the effect of precession. For
$\theta=\pi/3$, the two peaks at $f_\ast$ and $2 f_\ast$ have similar
strengths. The width of the peak, the lopsidedness, and the signal
power in between are numerical artifacts caused by the discrete Fourier
transform; they remain even when we input the unmodulated $\epsilon$
into \eref{emission:h0}. For $\theta=0.1$, a more realistic choice [cf. PSR B1828-11,
\citet{Link2003}], the peak at $2 f_\ast$ only reaches 28 (11) per cent of
the $f_\ast$ peak in the $+$ ($\times$) polarisation
(bottom panels of \fref{spectrum:axisym}). When the total signal power
is distributed into two peaks of similar height, coincidence
experiments at two frequencies become possible.

\begin{figure}
  \includegraphics[width=84mm, keepaspectratio]{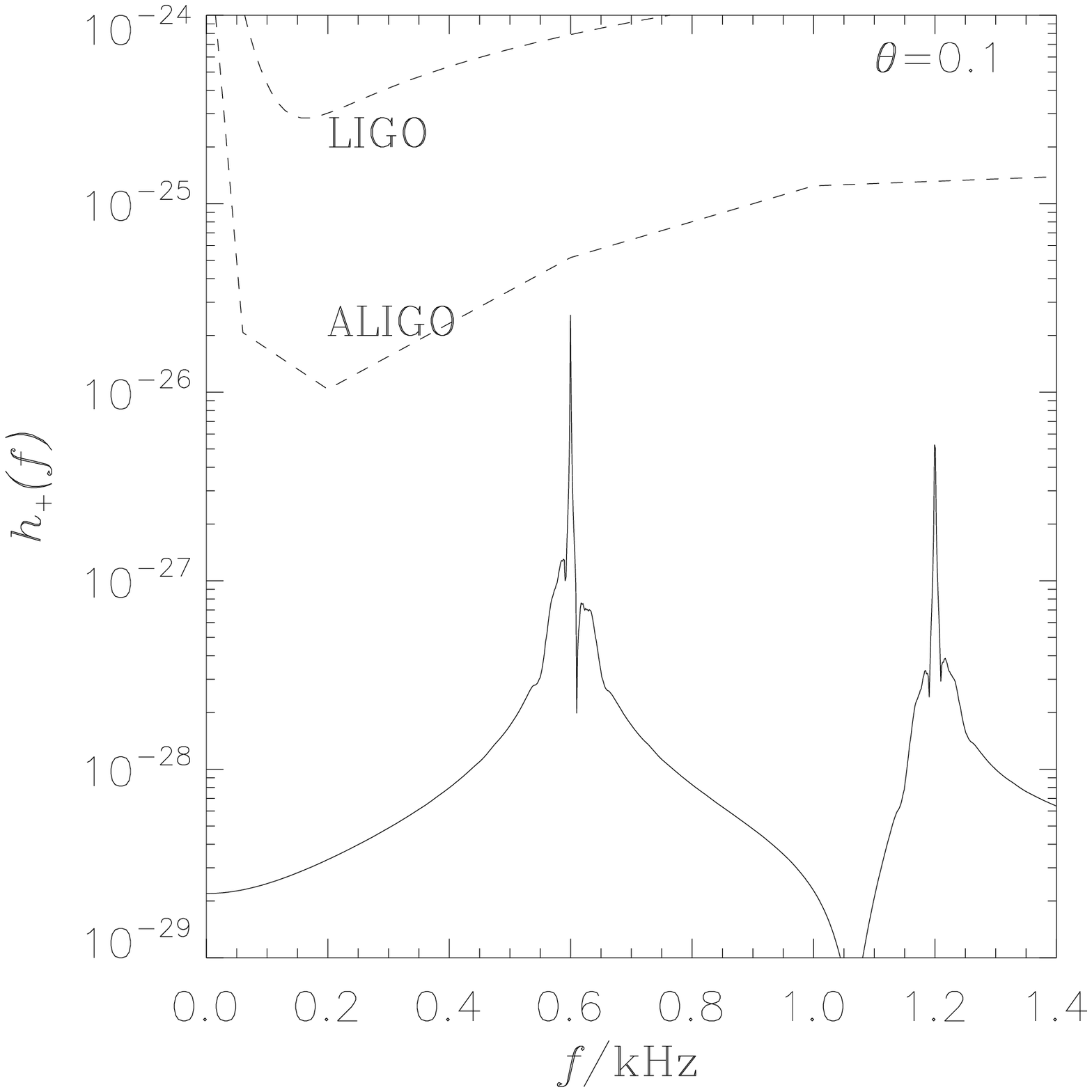}
  \caption{Fourier transform of $h_+(t)$, for a
    neutron star ($M_a=1.4 M_c$) at a distance $D=10$ kpc, with wobble
    angle $\theta=0.1$ and inclination $i=\pi/3$ (cf \fref{spectrum:axisym}). The
    amplitude of the oscillations in $\epsilon$ has been 
    increased artificially by a factor 10 in order to bring out the
    sidebands. The nonzero width of the sidebands stems from numerical
    damping.} 
  \label{fig:spectrum:axisym_sidebands}
\end{figure}

The acoustic and Alfv\'{e}nic sidebands produced by the global
oscillations are very hard to see in \fref{spectrum:axisym}, due to
strong numerical damping. In a realistic neutron star, the Alfv\'{e}n
mode may be perpetually re-excited (e.g. by fluctuations in the
accretion torque). Little is known at present about the excitation
mechanism. To demonstrate the effect, however, we artificially
increase the oscillation amplitude ten-fold \change{(with respect to
  the \textsc{zeus-mp} output)} by subtracting the time
average from the signal, multiplying the remaining signal by ten, and
finally adding the time average again. The results appear in
\fref{spectrum:axisym_sidebands}. The
sidebands are clearly visible, separated by $\Delta f = 17$ Hz from
the main peak. The width of the sidebands is set by the damping rate.

In principle, a high resolution spectrum of the gravitational
wave signal allows us to measure $M_a$ and the surface magnetic field
from $\Delta f$ \citep{Payne06a, Vigelius08a}. A detailed analysis of
the global MHD oscillations, including the linear response to a
stochastic excitation, will be attempted in a forthcoming paper.


\subsection{Triaxial, nonvibrating mountain}
\label{sec:spectrum:triaxial}
\begin{figure*}
  \includegraphics[width=168mm, keepaspectratio]{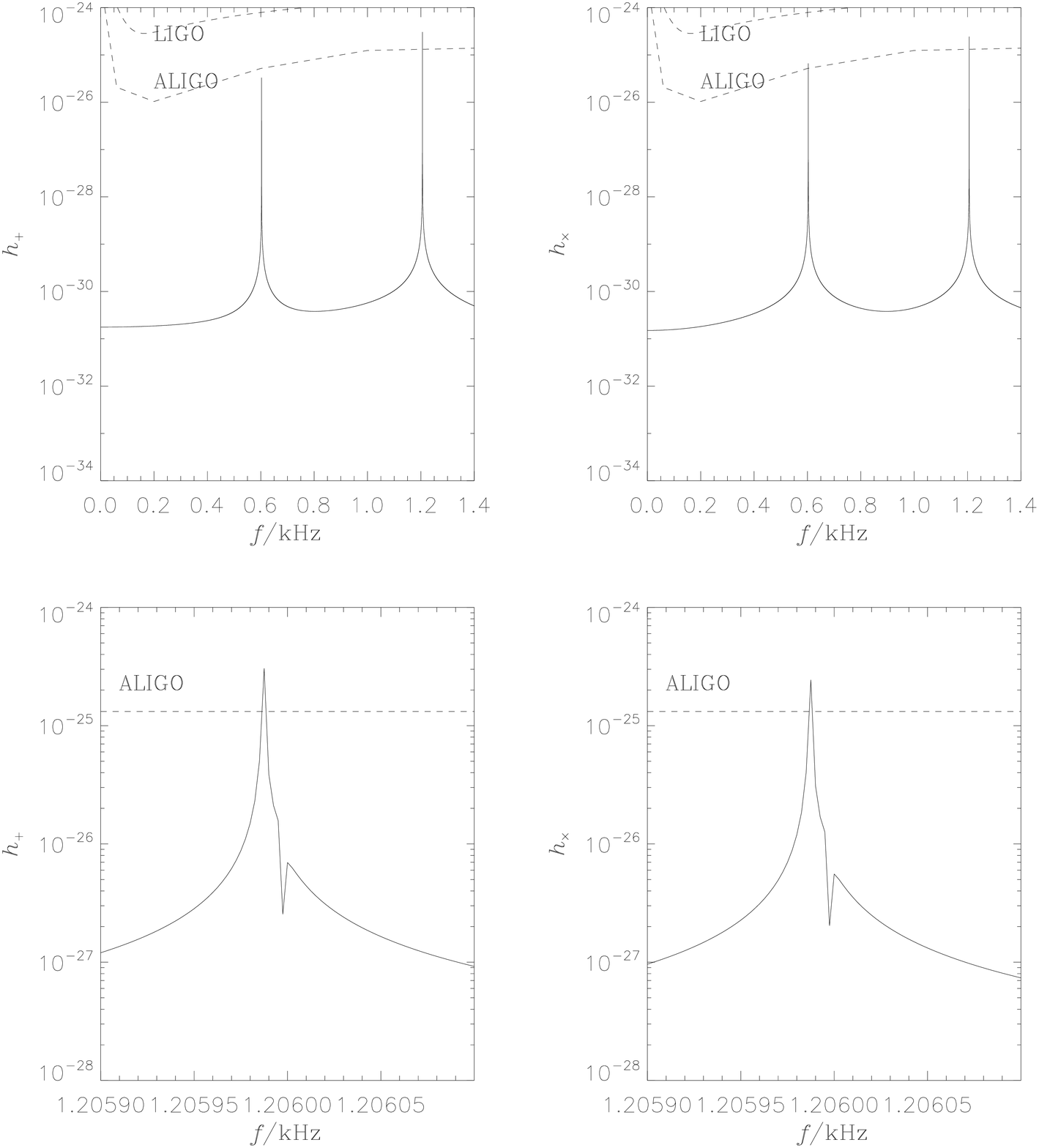}
  \caption{Fourier transform of the gravitational wave signal from a standard
    neutron star ($M_a=1.4 M_c$) with an nonaxisymmetric magnetic mountain at a distance
    $D=10$ kpc, with wobble angle $\theta=0.1$
    and inclination $i=\pi/3$. The left (right)
    panels show the discrete Fourier transform of the signals $h_+(t)$
    [$h_\times(t)$] in the plus and cross polarisations
    respectively (second order in $\theta$). The dashed curves
    display $h_\mathrm{th}$ of LIGO and advanced LIGO, assuming an
    observation time of $T_0=14$ d.  The
    lower panels zooms in on the peak at $2
    (\Omega_\mathrm{rot}+\Omega_\mathrm{prec})$, revealing the split
    peak due to triaxiality.}  
  \label{fig:spectrum:nonaxisym}
\end{figure*}

The Parker instability
experienced by an initially axisymmetric mountain saturates in a
slightly nonaxisymmetric state, with $B_\phi \sim B_p$ and
$Q_{12}/Q_{33} \sim 10^{-3}$, as discussed in see sections
\ref{sec:mountains:equilibrium} and
\ref{sec:mountains:hydro}. A third, unequal principal moment of
inertia introduces new features into the gravitational wave
spectrum. \citet{Zimmermann80} computed the waveform in a
small-wobble-angle expansion. \citet{vandenbroeck05} extended this
analysis up to second order in $\theta$, finding
\begin{equation}
  \label{eq:spectrum:nonaxi:hplus}
  h_+ (t)= \sum_{k=0}^1 \left[A^\mathrm{I}_{+,k} \cos
    (\Omega^\mathrm{I}_{2 k} t) + A^\mathrm{II}_{+,k} \cos
    (\Omega^\mathrm{II}_{2 k} t) \right]
\end{equation}
and
\begin{equation}
  \label{eq:spectrum:nonaxi:hcross}
  h_\times (t)= \sum_{k=0}^1 \left[A^\mathrm{I}_{\times,k} \sin
    (\Omega^\mathrm{I}_{2 k} t) + A^\mathrm{II}_{\times,k} \sin
    (\Omega^\mathrm{II}_{2 k} t) \right],
\end{equation}
with $\Omega^\mathrm{I}_{2 k}=2 \Omega_\mathrm{rot}+2 k
\Omega_\mathrm{prec}$ and $\Omega^\mathrm{II}_{2
  k}=\Omega_\mathrm{rot}+(1+ 2 k) \Omega_\mathrm{prec}$, where the
fundamental angular frequencies are
\begin{equation}
  \label{eq:spectrum:omegaprec}
  \Omega_\mathrm{prec} = \frac{\pi b}{2 K(m)} \left[
    \frac{(I_3 - I_2)(I_3-I_1)}{I_1 I_2} \right]^{1/2}
\end{equation}
and
\begin{equation}
  \label{eq:spectrum:omegarot}
  \Omega_\mathrm{rot}= \frac{J}{I_1}-\left[ 1+\frac{\mathrm{i}}{\pi}
    \frac{\vartheta_4' (\mathrm{i}\pi \alpha)}{\vartheta_4(\mathrm{i}\pi
      \alpha)} \right] \Omega_\mathrm{prec}.
\end{equation}
In \eeref{spectrum:omegaprec} and \eeref{spectrum:omegarot}, $I_1$,
$I_2$, and $I_3$ denote the principal moments of inertia,
$J=(I_1^2 \Omega_1^2 + I_2^2 \Omega_2^2 +
I_3^2 \Omega_3^2)^{1/2}$ is the (conserved) total angular momentum, and
$\mathbf{\Omega}=(\Omega_1, \Omega_2, \Omega_3)$ is the angular velocity
vector. At $t=0$, we have $\mathbf{\Omega}=(a, 0, b)$ without loss of
generality, such that $a/b \approx \theta$ for small wobble angles.
For consistency with \citet{vandenbroeck05}, the principal moments of
inertia are ordered such 
that $I_1 \le I_2 \le I_3$ and $J^2 \ge 2 E I_2$, where $2 E= I_1
\Omega_1^2 + I_2 \Omega_2^2 + I_3 \Omega_3^2$ is the (conserved) total
kinetic energy. The parameter $m$ is defined by
\begin{equation}
  m=\frac{(I_2-I_1) I_1 a^2}{(I_3-I_2) I_3 b^2},
\end{equation}
$K(m)$ is the complete elliptic integral of the first
kind, $\vartheta_4$ is the fourth Jacobi theta function, with
nome $q=\exp [-\pi K(1-m)/K(m)]$, and $\alpha$ is a solution of
\begin{equation}
  \mathrm{sn}[2 \mathrm{i} \alpha K(m), m] = \frac{\mathrm{i}I_3 b}{I_1 a},
\end{equation}
where sn is the (doubly periodic) Jacobi elliptic function. 

The amplitudes $A_+$ and $A_\times$ for the plus and cross
polarisations take the form \citep{vandenbroeck05}
\begin{equation}
  \label{eq:spectrum:amp1}
  A^\mathrm{I}_{+,0}=-\frac{2}{D} b^2 (1+\cos^2 i) (I_2-I_1),
\end{equation}
\begin{equation}
  A^\mathrm{I}_{+,1}=\frac{2}{D} b^2 [I_3-(I_1+I_2)/2] \gamma^2 (1+\cos^2 i),
\end{equation}
\begin{equation}
  A^\mathrm{I}_{\times,0}=-\frac{4}{D} b^2 (I_2-I_1) \cos i,
\end{equation}
\begin{equation}
  A^\mathrm{I}_{\times,1}=\frac{4}{D} b^2 [I_3-(I_1+I_2)/2] \gamma^2  \cos i,
\end{equation}
\begin{equation}
  A^\mathrm{II}_{+,0}=\frac{1}{D} b^2 [I_3-(I_1+I_2)/2] \gamma  \sin (2 i),
\end{equation}
\begin{equation}
  A^\mathrm{II}_{+,1}=0,
\end{equation}
\begin{equation}
  A^\mathrm{II}_{\times,0}=\frac{2}{D} b^2 [I_3-(I_1+I_2)/2] \gamma \sin i,
\end{equation}
and finally
\begin{equation}
  \label{eq:spectrum:amp6}
  A^\mathrm{II}_{\times,1}=0.
\end{equation}
where we omit the factor $G/c^4$ to avoid confusion.
In \eeref{spectrum:amp1}--\eeref{spectrum:amp6}, $i$ denotes the
inclination angle, $D$ is the distance to the
star, and $\gamma \ll 1$ is the expansion parameter
\begin{equation}
  \gamma=\frac{I_1 a}{I_3 b}.
\end{equation}

Equations \eeref{spectrum:nonaxi:hplus} and
\eeref{spectrum:nonaxi:hcross} point to the existence of three
distinct lines in the gravitational wave spectrum. Line I, at $2
\Omega_\mathrm{rot}$, stems from the departure from axisymmetry
[cf. the discussion in \citet{vandenbroeck05}]. A body rotating around
one of its  principal axes of inertia looks identical after
half a period. Line II, at $\Omega_\mathrm{rot}+\Omega_\mathrm{prec}$, results from the 
free precession of a nearly axisymmetric ($I_1 \rightarrow I_2$)
object in the small-wobble-angle approximation
\citep{Zimmermann80}. Line III, appearing as a sidelobe to line I at $\Omega^\mathrm{I}_0+2
\Omega_\mathrm{prec}$, results from the second order expansion of the
precession. For a biaxial star,
lines I and III coincide at the frequency
$\Omega_\mathrm{rot}+\Omega_\mathrm{prec}=2 \pi f_\ast$, consistent with
\eref{emission:signal}. Nonaxisymmetry then separates line I, which is
shifted to a lower frequency and the
second-order line III, which remains at $2\pi f_\ast$.

We compute the gravitational wave spectrum for the triaxial equilibrium
(\fref{mountains}b) from $Q_{ij}$ at $t=10 \tau_\mathrm{A}$. In order to
rescale $Q_{ij}$ to a realistic curvature (see
\sref{mountains:curvature}), we employ the relation $\epsilon =
Q_{33}/2 I_3 \propto R_\ast^2$ \citep{Melatos05}, which implies
$Q_{33} \propto M_\ast R_\ast^4$, and hence $Q_{ij} \propto M_\ast
R_\ast^4$ consistent with 
\citet{Vigelius08a} \footnote{By applying the 
same scaling to all components $Q_{ij}$, we
effectively assume that the three-dimensional structure of the 
mountain scales homologously with $R_\ast$. We find empirically that
this assumption holds for axisymmetric \citep{Payne04, Payne07} and
nonaxisymmetric configurations [see \citet{Vigelius08a}, and section 4.6].} .
The (upscaled) principal moments of inertia evaluate to $[I_1/\mathrm{Tr}(I)]-1 =
-3.9 \times 10^{-6}$, $[I_2/\mathrm{Tr}(I)]-1= -1.8 \times 10^{-6}$, and
$[I_3/\mathrm{Tr}(I)]-1=5.8 \times 10^{-6}$, where $\mathrm{Tr}(I)=3.3
\times 10^{45}$ g cm$^2$ is the trace of the moment of inertia
tensor. We also consider a star with $D=10$ kpc, $\Omega/2 \pi= 600$
Hz, and $\theta \approx a/b = 0.1$ rad. This particular choice
results in $\alpha=0.93$, $\Omega_\mathrm{prec}=0.032$ rad s$^{-1}$, and
$\Omega_\mathrm{rot}=3.7 \times 10^3$ rad s$^{-1}$.

\fref{spectrum:nonaxisym} depicts the spectrum (solid curves) emitted
by the above star, along with $h_\mathrm{th}$ for Initial LIGO and
Advanced LIGO (dashed curves), assuming an observation time of
$T_0=14$ d. Again, all the peaks in this mock spectrum are $\delta$
functions in reality; their width stems from the finite resolution of the discrete Fourier
transform. The peaks at $\Omega^\mathrm{I}_0$ and $\Omega^\mathrm{II}_0$ are
clearly visible in both polarisations. Neither surpasses the noise
floor unless the tunability of Advanced LIGO is exploited. A larger
wobble angle improves matters. The size of $\theta$ is controlled
be extraneous factors, such as the angle $\mathbf\Omega$ makes with the
accretion disk \citep{Lai07}, the accretion history,
the crystallisation history of the crust \citep{Melatos00, Melatos05}, and
dissipative processes in the superfluid interior \citep{Cutler87}, none of
which are well understood. However, if deeply modulated, persistent
X-ray pulsations emanate from a hot spot on the stellar surface at the
magnetic poles, then $\theta$ must be appreciable \citep{chung08}.

The bottom panels of \fref{spectrum:nonaxisym} zoom in on the peaks at
$\Omega^\mathrm{II}_0$, revealing the second-order peak at
$\Omega^\mathrm{II}_2$. We measure
the amplitudes to be in the ratios 
$|A^\mathrm{I}_{+,0}/A^\mathrm{I}_{+,1}| \approx
|A^\mathrm{I}_{\times,0}/A^\mathrm{I}_{\times,1}| \approx
25$. This shows clearly how triaxiality shifts the main peak
to a lower frequency ($2 \Omega_\mathrm{rot}$), while only a small, second-order peak
remains at  $2 (\Omega_\mathrm{rot}+\Omega_\mathrm{prec})$, as for a
biaxial star. The peaks are separated by $\Delta \Omega/2 \pi =
\Omega_\mathrm{prec}/\pi = 0.01$ Hz.

Can the extra, ``triaxial'' peak be exploited to facilitate detection,
i.e. to increase the signal-to-noise ratio $d$? The generalisation of
the results of \sref{gw:detectability} is straightforward, as long as
we assume that the detector can resolve all three spectral
lines. After averaging over sky position, inclination, and
polarisation, we find
\begin{equation}
  \left< d^2 \right> \simeq \left< d_\mathrm{I}^2 \right>+\left< d_\mathrm{II}^2 \right>
  + \left< d_\mathrm{III}^2 \right>
\end{equation}
with
\begin{equation}
  \left< d_\mathrm{I}^2 \right>=\frac{16 T_0 b^4}{25 S_h(f_\mathrm{I}) D^2} (I_2-I_1)^2,
\end{equation}
\begin{equation}
  \left< d_\mathrm{II}^2 \right>=
  \frac{16 T_0 b^4}{25 S_h(f_\mathrm{II}) D^2} [I_3-(I_1+I_2)/2]^2 \gamma^2,
\end{equation}
\begin{equation}
  \left< d_\mathrm{III}^2 \right>=
  \frac{16 T_0 b^4}{25  S_h(f_\mathrm{III}) D^2} [I_3-(I_1+I_2)/2]^2 \gamma^4,
\end{equation}
with $f_\mathrm{I,II}=\Omega^\mathrm{I, II}_0/(2 \pi)$ and
$f_\mathrm{III}=\Omega^\mathrm{II}_2$.

If the canonical star above ($M_a=10^{-5} M_\odot$, $D=10$ kpc, and
$\theta=0.1$) is observed with Initial LIGO, we expect $\left< d_\mathrm{I}^2 \right>
= 1.6$, $\left< d_\mathrm{II}^2 \right> = 1.1$, and
$\left< d_\mathrm{III}^2 \right> = 0.003$. Importantly, if one does not account for
the frequency shift of line II when searching the LIGO data, one picks up only the strongly
attenuated line III and hence only half of the total signal-to-noise
ratio.

Once an initial detection is made, it is possible
to extract the inclination, precession angle, deviation
from axisymmetry, and  oblateness parameter $1-(I_1+I_2)/2 I_3$
\citep{vandenbroeck05} from the gravitational wave spectrum. Firstly, $i$ can
be found easily from the amplitude ratio of the two polarisations
in any spectral line. Next, the amplitude ratio of line II and line III can
(knowing $i$) be used to infer $I_1 a/I_3 b$. The ratio of lines I and II
is proportional to $I_3 b(I_2-I_1)/I_1 a[I_3-(I_2-I_1)/2]$. The
line frequencies tell us $\Omega_\mathrm{rot}$ and
$\Omega_\mathrm{prec}$. Using the approximate relation
$\Omega_\mathrm{prec}/\Omega_\mathrm{rot} \approx \pi
[I_3-(I_1+I_2)/2]/2 K(m) I_3$, one can finally derive the oblateness
parameter $[I_3-(I_1+I_2)/2]/I_3$.

\subsection{Precession amplitude}
At present, the wobble angle $\theta$ is partly constrained by theory
and observation. We know that it cannot be zero exactly, because the
fluctuating magnetospheric accretion torque has nonzero
components perpendicular to $\mathbf{\Omega}$ \citep{Lai99}, which
give $\theta \ne 0$ for finite damping. But how small is the
steady-state $\theta$? If the magnetic axis $\bmu$ 
is aligned with $\bmath{\Omega}$, due to viscous dissipation, before the crust
crystallises, then $\bmu$, $\bmath{\Omega}$, and $\bmath{e}_3$ are all
aligned. In this case, we see neither precession nor X-ray pulsations
before accretion begins (provided the pulsations stem from
a hot spot at the magnetic pole). However, a subsequently accreted mountain with a
small nonaxisymmetry emits gravitational radiation at $f=2
\Omega_\mathrm{rot}$ (line I), but still no X-ray radiation. On the
other hand, if the crust crystallises before $\bmath{\Omega}$ aligns
with $\bmath{\mu}$, we see precession, pulsation, and gravitational radiation
in all three lines \citep{Melatos00, Payne06a}.

Gravitational-wave back-reaction damps the wobble on
a time-scale $\tau \sim 10^5 \mathrm{yr} (10^{-7}/\epsilon)^2
(\mathrm{kHz}/f_\ast)^4$ \citep{Cutler01}, provided the star can be
treated as a fluid body with an elastic crust and the
precession is torque-free. On the other hand, a neutron star with an
accreted mountain forms a prolate spheroid. In this case, internal dissipation
increases the wobble angle ($\theta \rightarrow \pi/2$) [cf. the
discussion in \citet{Cutler01}]. \citet{Alpar88} examined the coupling
of the superfluid interior and the stellar crust, finding that
$\theta$ approaches $\pi/2$ over $\sim 10^4$ free precession
periods.

It has been postulated that precession explains the slow variation in
pulse-arrival-time residuals observed in accreting neutron stars
(e.g. Her X-1) and isolated pulsars (e.g. PSR B0959-54 and PSR
B1828-11) [cf. the discussion in \citet{Akgun06}]. \citet{chung08}
modelled the X-ray flux modulations observed in XTE J1814$-$338. They found
$\epsilon \le 10^{-9}$ and $0\degr \le \theta \le
5\degr$ if the star precesses and $\epsilon \cos \theta \le 10^{-10}$
if not. \citet{Akgun06}
modelled the radio pulses from PSR B1828$-$11, finding that
they can be matched by a precessing, triaxial star, whose
degree of nonaxisymmetry is small and whose shape is
prolate. The wobble angle of this isolated pulsar is $\theta \sim
3\degr$ \citep{Link2003, Stairs00}.

\section{Discussion}
\label{sec:discussion}
In this article, we give improved estimates for the strength and
spectrum of gravitational waves from accreting millisecond
pulsars. For the first time, we include the effects of nonaxisymmetry
and resistive relaxation on the gravitational wave
signal. Furthermore, we justify the curvature downscaling introduced in
previous work \citep{Payne04, Payne07, Vigelius08a} and achieve
self-consistent mountain configurations with $M_a \sim 10^{-3}
M_\odot$, ten times larger than constructed previously, by growing
them from scratch by injection. \change{We cannot find any evidence
  for a growing, axisymmetric instability even at these $M_a$.}

Taken at face value, Figs. \ref{fig:emission:snr} and \ref{fig:emission:detectability}
imply that mountains with $(f_\ast/0.3\,\mathrm{kHz})^2 (M_a/10^{-5}
M_\odot) \ga 1$ are detectable in principle even with Initial LIGO. With Advanced
LIGO, chances are even better over the whole $0.1 - 1$ kHz band
for $M_a \ga 10^{-6} M_\odot$. While this encouraging result was
foreshadowed by \citet{Melatos05}, this paper includes for the first
time the dynamics of nonaxisymmetry and resistive
relaxation, which act to weaken the signal -- yet still the prospects
remain bright. In principle, the noise floor of Advanced LIGO can be
lowered even further by using a narrowband configuration exploiting a
squeezed vacuum \citep{Buonanno04}, improving the sensitivity by as
much as threefold for $f \la 400$ Hz. A successful detection will let
us test the reciprocal dependence of the gravitational wave amplitude
on the magnetic dipole moment predicted for magnetic mountains by
\citet{Melatos05}.

Of course, magnetic mountains have not been detected during recent
searches for low-mass X-ray binaries in the S5 LIGO data
\citep{LIGO06, Watts08}. Moreover, for many LMXBs, the quadrupole
moment predicted by \fref{mountain:growth_ellipticity} is too large to
be consistent with the measured spin frequency, because the
accretion-driven recycling process is ``stalled'' by the
gravitational-wave spin-down torque \citep{Bildsten98, Chakrabarty03}.
This suggests that other physical
processes exist, unaccounted for so far, which reduce the
mountain. The circle of candidates has shrunk significantly since the exclusion of
resistive relaxation\footnote{It is possible that the electrical
  resistivity is substantially higher than the contribution from
  electron-phonon scattering \citep{Cumming04} or that the Hall effect
  plays a dominant relaxing role \citep{Pons07}. These possibilities
  will be pursued in future investigations.}\citep{Vigelius08b}, but there are others,
principally hydrodynamic sinking. In all the modelling to date, we
assume that the mountain sits on a hard surface. Realistically, the
neutron star crust is not impenetrable, and part of the
mountain sinks into it \citep{Konar97, Choudhuri02,
  Payne04}\footnote{\citet{Zhang06} took into account the contraction
  of the magnetosphere 
during magnetic burial. Ultimately, the equatorial belt is pushed
into the crust when the accretion disk touches the surface of the
star. In agreement with \citet{Payne05}, one needs $M_a \ga 10^{-5}
M_\odot$ to significantly reduce the magnetic dipole moment.
The ``bottom'' magnetic field set by magnetospheric contraction
depends on the accretion rate.}. An attempt to treat
sinking by growing mountains on top of a soft crust, modelled by a
polytropic equation of state, will be presented elsewhere [Wette et
al. (in preparation)]. A more sophisticated model would include the
realistic, stratified nuclear composition of the crust and its evolution
in response to pycnonuclear reactions \citep{Brown98,
  Ushomirsky00}. The Coriolis force may also push the mountain to
wander across the surface, especially in the fastest spinning 
LMXBs, although the enhanced magnetic field in the equatorial belt
opposes the wandering \citep{Payne06b}.



\change{For simplicity, we assume an isothermal equation of
  state throughout this article. During the late stages of accretion
  ($M_\mathrm{a} \ga 10^{-3}   M_\odot$), however, the magnetic mountain mass is comparable to the mass of
  the neutron star crust and the equilibrium exhibits a wide range of
  density and temperature. Pycnonuclear reactions in the deep regions
  ($\rho \ga 10^{12}$ g cm$^{-3}$) feed thermal energy into an
  adiabatic mountain. The assumption of
  isothermality breaks down and a realistic equation of state for
  non-catalyzed matter is required \citep{Haensel90a}. In particular,
  the accreted material is expected to solidify at densities $\ga 10^8$ g
  cm$^{-3}$ \citep{Haensel90b} and the crust needs to be modelled as
  an elastic solid \citep{Ushomirsky00}. The effect of
  a realistic equation of state is subject of current work and the results
will be presented elsewhere.} 

Finally, we note that the full (discrete and continuous) spectrum of
global MHD mountain oscillations contains valuable information about the
structure of the star, e.g. the strength of the surface magnetic
field. In principle, this spectral information will be accessible by
third-generation gravitational-wave interferometers with improved
sensitivity and frequency resolution. As a first step, we will
compute the continuous part of the MHD spectrum in a forthcoming
paper.

\bibliography{gravwave.bib}

\appendix

\section{Growing mountains with \textsc{zeus-mp}}
\label{sec:app:grow}
In this appendix, we describe one way to grow a mountain with $M_a \gg
M_c$, using the ideal-MHD code \textsc{zeus-mp}
\citep{Hayes06} extended to include Ohmic dissipation \citep{Vigelius08b}.

We start with an isothermal atmosphere (initial mass $\ll M_c$)
resting on a hard surface at $r=R_\ast$ threaded by a dipolar magnetic
field. We then inject matter quasistatically into the atmosphere from
below (at $r=R_\ast$) along a polar subset of the magnetic field
lines, imitating  disk-fed accretion. We artificially boost the mass
flux so as to grow the mountain in a reasonable time, albeit slowly
compared to the hydromagnetic Alfv\'{e}n time-scale. This is achieved
by boosting the injection mass density $\rho_0$ while keeping the
injection speed $v_0$ below the gravitational escape speed
(i.e. escape from the simulation volume).

Mass is added at $r=R_\ast$ at a rate $\dot{M} (\theta) = \rho_0 v_0
\exp (-b \sin^2 \theta)$, where $b$ is the polar cap radius scaled to
$R_\ast$ \citep{Payne04}. By injecting at $r=R_\ast$ (from ``below'')
instead of $r=R_m$ (from ``above''), we exploit the flux-freezing
property of ideal MHD to avoid having to 
locate the changing intersection point of the flux surface
$\Psi=\Psi_0$ with $r=R_m$; cf. the bootstrapping algorithm in
\citet{Payne07}. We find empirically that a good compromise between
shortening the run time while maintaining quasistatic injection is
$\rho_0=5.1 \times 10^{10}$ g cm$^{-3}$ and $v_0 = 10^{4}$ cm
s$^{-1}$.

Although the magnitude of $\bmath{v_0}$ is constant, its direction
must be parallel to the dipolar magnetic field everywhere at
$r=R_\ast$, in order to respect the flux-freezing constraint. The
$\theta=\pi/2$ surface is reflecting [\texttt{ojb.nojs(1)= 5}],
with normal magnetic field, which translates to
$\mathbf{v}_\perp=\mathbf{B}_\parallel=0$. The line $\theta=0$
is also reflecting [\texttt{ijb.nijs(1)= -1}], with tangential magnetic field
($\mathbf{v}_\perp=\mathbf{B}_\perp=0$). The outer $r$ surface is
\texttt{outflow} [cf. section 3.2 in \citet{Vigelius08a}].

\begin{figure}
  \includegraphics[width=84mm,keepaspectratio]{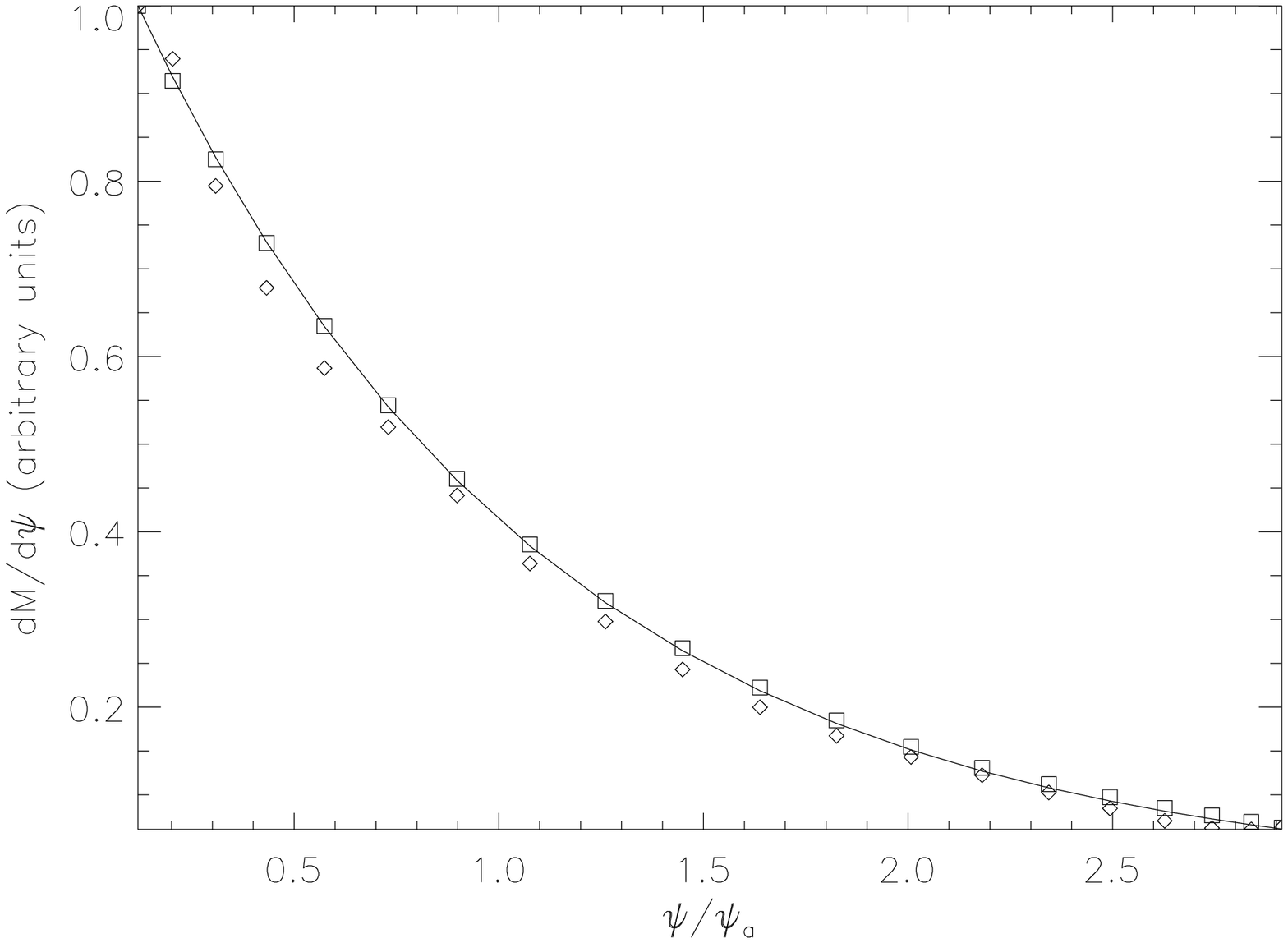}
  \caption{$dM/d\psi$ in arbitrary units as a function of $\psi/\psi_a$
  for the Grad-Shafranov equilibrium (diamonds) and the injection
  simulation (square). For comparison, we also include the theoretical
  value $dM/d\psi=\exp(-\psi/\psi_a)$ (solid curve). The snapshot is
  taken at $t=2 \tau_\mathrm{A}$, although the mass-flux ratio remains
  constant with time, of course.}
  \label{fig:implementation:dmdpsi}
\end{figure}

We first verify that the growing mountain upholds the correct
mass/flux ratio. \fref{implementation:dmdpsi} displays $dM/d\psi$ for
the injection simulation (squares), the Grad-Shafranov equilibrium
(diamonds), and the analytic distribution $dM/d\psi=\exp(-\psi/\psi_a$)
\citep{Payne04}. All match to better than one per cent. While the
snapshot is taken at $t=2 \tau_\mathrm{A}$ (which translates to
$M_a=0.46 \times 10^{-4} M_\odot$), we verify that $dM/d\psi$ remains unchanged
with time.

\begin{figure}
  \includegraphics[width=84mm,keepaspectratio]{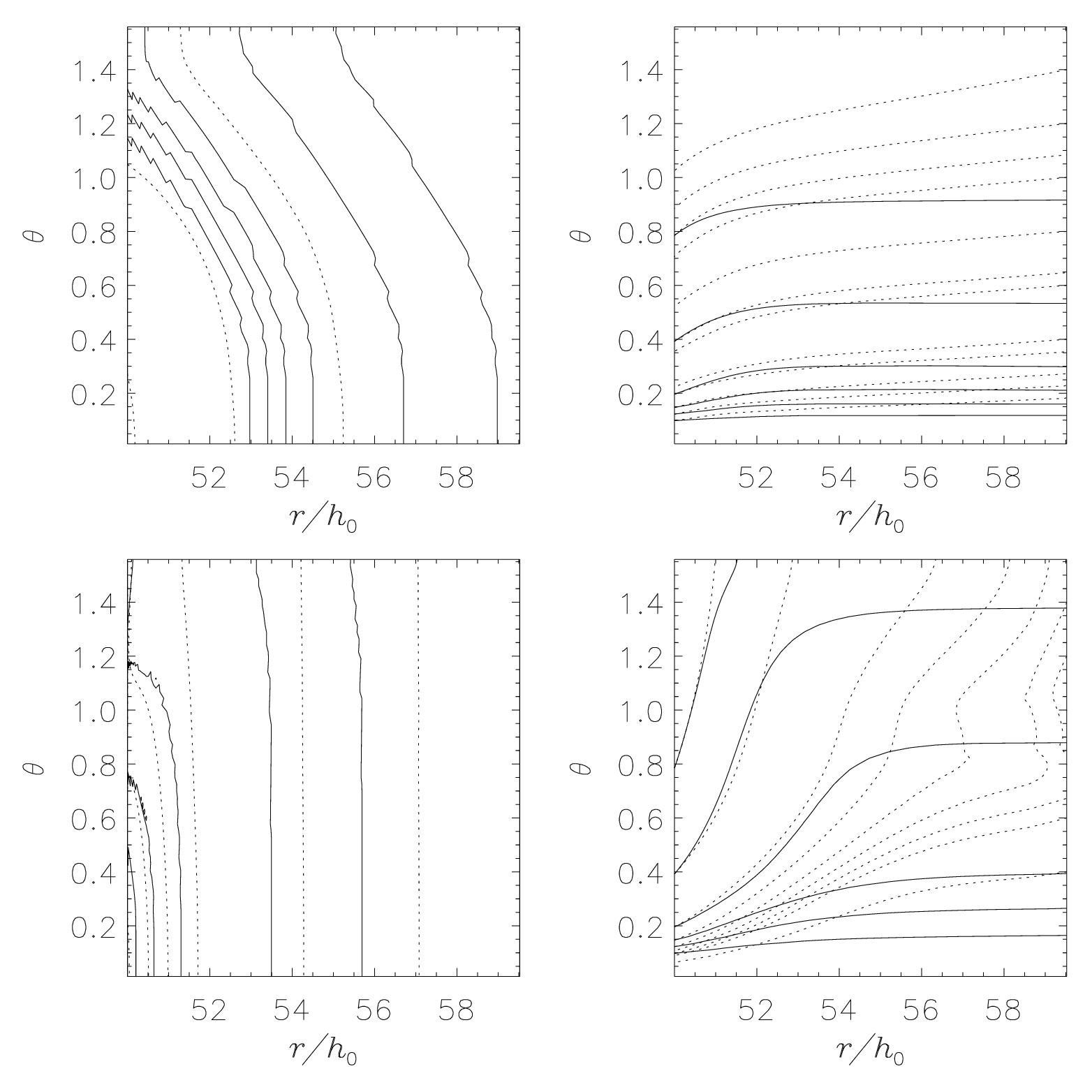}
  \caption{Meridional section of density contours (left) and magnetic
    field lines (right) for a grown mountain (dashed curve) and a
    Grad-Shafranov equilibrium (solid curve) with $M_a=0.1 M_c$ (upper
    panels) and $M_a=M_c$ (lower panels). While the agreement in the
    density is good, discrepancies in the magnetic field arise due to
    the different boundary conditions at $r=R_m$ (see text).}
  \label{fig:implementation:grow_overlay}
\end{figure}

By way of verification, we compare a grown mountain to a
Grad-Shafranov equilibrium for  $M_a=1.2 \times 10^{-4}$ in
\fref{implementation:grow_overlay}. The left panels display the
same density contours for the grown mountain (dashed curves) and the
equilibrium (solid curves), for $M_a=0.1 M_c$ (top panels) and
$M_a=M_c$ (bottom panels). The density contours in the region close to
$0 \le \tilde{x}=(r-R_\ast)/h_0 \le 2$ match reasonably well, with a deviation of less than one
per cent radially and five per cent laterally. While the magnetic
field lines match close to the stellar surface, there is a
considerable discrepancy in the (less important) outer regions, where $B$ is weak (and
indeed $\rho$ is small). This
stems from the different implementation of the outer boundary
condition at $r=R_m$. The \texttt{outflow} boundary condition in \textsc{zeus-mp}
enforces a vanishing gradient of $\mathbf{B}$ at $r=R_m$, while
\citet{Payne04} enforce $\partial \psi/\partial r=0$, thereby imposing
the additional constraint $B_\theta=0$. The exact form of the outer
boundary condition depends on the interaction of the magnetosphere
with the accretion disk and is poorly known. We are not concerned with the
details and just note that the magnetic field
decays $\propto r^{-3}$, such that it is several orders of magnitude
lower than at the surface.

\end{document}